\documentclass[smallextended]{svjour3}

\usepackage{graphicx}
\usepackage{amsmath,amssymb,amsfonts}
\usepackage{bm}
\usepackage{booktabs}
\usepackage[table]{xcolor}
\usepackage{siunitx}
\usepackage[numbers,sort&compress]{natbib}
\usepackage{subcaption}
\usepackage{textcomp}
\usepackage{hyperref}
\hypersetup{
colorlinks=true, citecolor=blue, linkcolor=blue, urlcolor=blue
}

\graphicspath{{figures/}}
\definecolor{tablehead}{gray}{0.88}
\definecolor{tablerow}{gray}{0.95}
\newcommand{\mathsfbi}[1]{\bm{\mathsf{#1}}}
\smartqed
\journalname{Theoretical and Computational Fluid Dynamics}

\begin{document}

\title{Receptivity and Biorthogonal Decomposition in a Reacting Temporal Mixing Layer}
\titlerunning{Receptivity and biorthogonal decomposition}

\author{Sriram P. Kalathoor \and Joseph C. Oefelein}
\authorrunning{S. P. Kalathoor and J. C. Oefelein}

\institute{S. P. Kalathoor \at
Daniel Guggenheim School of Aerospace Engineering, Georgia Institute of
Technology, Atlanta, GA 30332, USA \\
\email{sriram2@gatech.edu}
\and
J. C. Oefelein \at Daniel Guggenheim School of Aerospace Engineering, Georgia Institute of Technology, Atlanta, GA 30332, USA}

\date{Received: date / Accepted: date}

\maketitle

\begin{abstract}
We examine receptivity and biorthogonal decomposition in a reacting temporal mixing layer using direct and adjoint eigenmodes of a finite-thickness compressible linearized operator built from the mean reacting base state. The analysis focuses on the Kelvin--Helmholtz branch and asks how the reacting base state modifies the selected temporal instability, where localized forcing most efficiently excites it, and how strongly the associated modal family is represented in time-resolved planar simulation data. Receptivity maps are constructed for mass, momentum, thermal, and mixture-fraction forcing channels using an energy-weighted adjoint projection, with biorthogonality enforced by the corresponding direct--adjoint inner product. A complementary biorthogonal decomposition provides modal amplitudes and cumulative few-mode reconstructions at the fundamental streamwise wavenumber. The finite-thickness branch is interpreted against a compressible vortex-sheet reference built from the outer-stream states. The reacting layer supports an unstable finite-thickness Kelvin--Helmholtz family over low-to-moderate wavenumbers even though the discontinuous reference is essentially neutral. Mass forcing leads the raw localized receptivity maps, mixture-fraction forcing follows through composition-pressure coupling, and chemistry-weighted thermal forcing identifies the strongest thermochemical support of the same family. The results show how distributed reacting thermodynamics reorganize compressible shear-layer instability and how that reorganized branch remains embedded in the nonlinear flow.

\end{abstract}

\section{Introduction}
Receptivity analysis connects external or internal forcing to the amplitude of unstable modes through adjoint sensitivity, while biorthogonal decomposition provides the corresponding modal-amplitude framework for non-normal linear systems. For a temporal mixing layer, these two viewpoints are naturally complementary: the direct eigenproblem identifies the unstable branch and its cross-stream structure, the adjoint eigenproblem identifies where forcing most efficiently excites that branch, and the biorthogonal projection supplies a modal coordinate system in which the simulation data can be interpreted
\citep{BaroneLele2005,Schmid2007,TuminWangZhong2007,AlHasnineRussoTuminBrehm2023}.

The present study focuses on a reacting temporal mixing layer in which compressibility, thermochemistry, and finite-thickness shear all modify the classical Kelvin--Helmholtz instability. The paper therefore centers on how the reacting mean state reshapes the instability branch, its receptivity, and its modal imprint in the nonlinear data. The central objects are the mean reacting base state, the selected Kelvin--Helmholtz branch of the finite-thickness temporal operator, localized forcing maps for a small set of physically interpretable channels, and time-resolved planar modal amplitudes obtained directly from the simulation data. The compressible-mixing-layer context for this problem is set by the experimentally observed reduction in shear-layer growth with increasing convective Mach number \citep{PapamoschouRoshko1988}. That trend sits within a broader free-shear-layer literature in which density ratio, entrainment, and the persistence of large-scale organization were identified as central features of turbulent mixing layers
\citep{BrownRoshko1974,DimotakisBrown1976}. Chemical-reaction markers were then
used to expose the internal large-scale structure of turbulent mixing layers and wakes \citep{Breidenthal1981}, while simplified mixing-layer models clarified how mixing and reaction remain organized by the same dominant motions \citep{BroadwellBreidenthal1982}. For compressible layers, linear theory, experiment, DNS, and LES have shown that increasing compressibility suppresses growth, modifies entrainment, and reorganizes the large-scale vortical field without eliminating coherent Kelvin--Helmholtz-type structure altogether
\citep{SandhamReynolds1990,SandhamReynolds1991,VremanSandhamLuo1996,Sarkar1995,ElliottSamimy1990,ClemensMungal1995,FreundLeleMoin2000,FoysiSarkar2010,JahanbakhshiMadnia2016,MatsunoLele2020}.
Taken together, these studies establish the macroscopic effect of compressibility on shear-layer growth while also showing that the compressible layer retains coherent Kelvin--Helmholtz-type organization even as the instability and turbulence depart from incompressible expectations.

On the analytical side, adjoint receptivity methods provide a natural route for identifying where forcing most efficiently excites unstable compressible shear layers. For the compressible mixing layer, Barone and Lele \citep{BaroneLele2005} showed how direct and adjoint modes can be combined to map localized forcing sensitivity over wavenumber and cross-stream position. More broadly, non-normal stability theory and adjoint-based disturbance analysis have clarified how direct modes, adjoint modes, and disturbance-energy inner products together determine modal amplification and receptivity in compressible flows
\citep{Chu1965,Goldstein1983,Goldstein1985,Hill1995,Schmid2007,TairaEtAl2017}.
For free shear layers in particular, receptivity to acoustic or localized external excitation was already recognized in early theoretical studies \citep{Tam1978,Balsa1988}. Recent biorthogonal decompositions in high-speed flow problems have extended this viewpoint from sensitivity analysis to data interpretation by using direct/adjoint mode pairs as a modal coordinate system for complex disturbance fields \citep{TuminWangZhong2007,AlHasnineRussoTuminBrehm2023}.

The reacting temporal-layer setting adds a second layer of complexity because the instability problem is filtered through variable thermodynamic and transport coefficients that are themselves organized by the evolving mean composition and temperature fields. DNS studies of high-pressure reacting temporal mixing layers have shown that thermochemistry and variable transport materially alter the internal structure of the turbulent layer \citep{Bellan2017}. Earlier reacting-shear-layer experiments and simulations showed that heat release can reorganize large-scale structure, alter mixing, and modify turbulence spectra in free shear flows \citep{HermansonDimotakis1989,McmurtryRileyMetcalfe1989,PantanoSarkarWilliams2002,PantanoSarkarWilliams2003,MahleEtAl2007,KnausPantano2009,JahanbakhshiMadnia2018}. On the instability side, classical compressible vortex-sheet and inviscid mixing-layer analyses demonstrated how sensitively unstable branches depend on outer-state compressibility and thermodynamic closure \citep{Miles1958,Blumen1970,BlumenDrazinBillings1975,DrazinDavey1977,JacksonGrosch1989}. Reacting flame-sheet extensions showed that heat release can change mode families and, in idealized limits, even promote absolute instability \citep{JacksonGrosch1990}. Localized energy deposition has also been shown to be an effective receptivity pathway in compressible mixing layers \citep{WuTumin2019}, suggesting that the forcing hierarchy of a reacting layer may differ qualitatively from that of a non-reacting one. Any sustained departure from those idealized compressible limits in a finite-thickness reacting layer therefore has direct implications for how turbulent compressible reacting flows retain coherent shear-layer organization.

The present work builds on these threads by combining a finite-thickness reacting temporal operator, adjoint receptivity maps for mass, momentum, thermal, and mixture-fraction channels, and DNS-based biorthogonal modal projections within a single framework. The emphasis is on the physics of how reacting thermodynamic structure alters the Kelvin--Helmholtz branch, how that altered branch selects forcing channels, and how strongly it remains represented in the turbulent reacting flow field. The central question is what the branch reveals about coherent unsteadiness, forcing preference, and thermochemical control in turbulent compressible reacting mixing layers. The paper extends classical compressible mixing-layer receptivity theory into a reacting temporal setting, ranks the forcing pathways of the selected Kelvin--Helmholtz family, and uses the DNS-projected modal content to connect that extension back to the physics of a turbulent compressible reacting shear layer.

\section{Problem Formulation}
\subsection{Reacting temporal mixing layer and mean base state}
We consider a reacting compressible temporal mixing layer with conserved variables
\begin{equation}
\bm{Q} =
\left(
\rho,\,
\rho u,\,
\rho v,\,
\rho w,\,
\rho E,\,
\rho Z
\right)^{\mathsf{T}},
\end{equation}
where $\rho$ is the density, $(u,v,w)$ are the velocity components, $E$ is the total specific energy, and $Z$ is the mixture fraction. The corresponding primitive state is written as
\begin{equation}
\bm{q} =
\left(
\rho,\,
u,\,
v,\,
w,\,
T,\,
Z
\right)^{\mathsf{T}},
\end{equation}
with temperature $T$ closing the thermodynamic state through tabulated thermochemical relations.

The analysis is performed on time-resolved planar $x$--$y$ slices extracted at a fixed spanwise location from the three-dimensional temporal simulation. A mean base state is defined by streamwise averaging at each time followed by a time average over the retained planar data sequence. Denoting the streamwise average of a quantity $f$ by $\langle f \rangle_x$ and the time average by $\langle \cdot \rangle_t$, the Reynolds mean profile is
\begin{equation}
\bar{f}(y)=\left\langle \langle f \rangle_x \right\rangle_t,
\end{equation}
while Favre-averaged velocity, temperature, and mixture-fraction profiles are defined by
\begin{equation}
\tilde{f}(y)=\frac{\left\langle \langle \rho f \rangle_x \right\rangle_t}
{\left\langle \langle \rho \rangle_x \right\rangle_t}.
\end{equation}
The resulting base state is therefore
\begin{equation}
\bar{\bm{q}}(y)=
\left(
\bar{\rho},\,
\tilde{u},\,
\tilde{v},\,
\tilde{w},\,
\tilde{T},\,
\tilde{Z}
\right)^{\mathsf{T}}.
\end{equation}

For the discussion and plotting below, base-state quantities are nondimensionalized with lower-stream reference values and the outer-stream velocity difference,
\begin{equation}
\rho^*=\frac{\rho}{\rho_b},\qquad
u^*=\frac{u}{\Delta U},\qquad
T^*=\frac{T}{T_b},\qquad
R_{\mathrm{mix}}^*=\frac{R_{\mathrm{mix}}}{R_b},
\end{equation}
\begin{equation}
\mu^*=\frac{\mu}{\mu_b},\qquad
\kappa^*=\frac{\kappa}{\kappa_b},\qquad
D_Z^*=\frac{D_Z}{D_{Z,b}},\qquad
y^*=\frac{y-y_0}{\delta_{\omega,\mathrm{ref}}},
\end{equation}
where $\Delta U=U_t-U_b$, $y_0$ is the layer-centre reference position, and
\begin{equation}
\delta_{\omega,\mathrm{ref}}
=\frac{U_t-U_b}{\max_y |\mathrm{d}\tilde{u}/\mathrm{d}y|}
\end{equation}
is the reference vorticity thickness of the mean reacting base state.

Thermodynamic and transport quantities along this base state are sampled from a precomputed chemistry table on the $(Z,T)$ manifold. In particular, we retain
\begin{equation}
R_{\mathrm{mix}}(y),\quad
\gamma(y),\quad
c_p(y),\quad
c_v(y),\quad
\mu(y),\quad
\kappa(y),\quad
D_Z(y),
\end{equation}
together with pressure derivatives with respect to density, temperature, and mixture fraction evaluated along the base state. These profiles define the variable-coefficient temporal operator used in the modal analysis.

\subsection{Linearized temporal eigenproblem}
Perturbations are introduced through
\begin{equation}
\bm{q}'(x,y,t)=\bm{q}(x,y,t)-\bar{\bm{q}}(y),
\end{equation}
and the reduced forced linearized system is written as
\begin{equation}
\partial_t\bm{q}' + \mathcal{L}\bm{q}' = \bm{f}.
\label{eq:forced_system}
\end{equation}
For the present reacting model, $\mathcal{L}$ contains advection by the mean streamwise velocity, cross-stream coupling through the base gradients $\rho_y$, $U_y$, $T_y$, and $Z_y$, the local thermodynamic closure
\begin{equation}
p' = p_\rho \rho' + p_T T' + p_Z Z',
\end{equation}
and reduced diffusive terms based on
\begin{equation}
\nu(y)=\frac{\mu(y)}{\rho(y)},\qquad
\alpha(y)=\frac{\kappa(y)}{\rho(y)c_v(y)},\qquad
D_Z(y)=D_{\mathrm{mix}}(y).
\end{equation}
Finite-rate chemistry enters this reduced eigenproblem through the reacting mean profiles and the table-derived thermodynamic and transport derivatives; tabulated thermal and composition source-rate magnitudes are used subsequently as support functions for chemistry-weighted receptivity maps. The explicit primitive-variable system is
\begin{align}
\partial_t \rho' &=
-\tilde{u}\,\partial_x \rho'
-\rho\,\partial_x u'
-\rho\,\partial_y v'
-v'\rho_y, \\
\partial_t u' &=
-\tilde{u}\,\partial_x u'
-U_y v'
-\rho^{-1}\partial_x p'
\,+\,\nu \nabla^2 u', \\
\partial_t v' &=
-\tilde{u}\,\partial_x v'
-\rho^{-1}\partial_y p'
\,+\,\nu \nabla^2 v', \\
\partial_t w' &=
-\tilde{u}\,\partial_x w'
\,+\,\nu \nabla^2 w', \\
\partial_t T' &=
-\tilde{u}\,\partial_x T'
-T_y v'
-(\gamma-1)T(\partial_x u' + \partial_y v')
\,+\,\alpha \nabla^2 T', \\
\partial_t Z' &=
-\tilde{u}\,\partial_x Z'
-Z_y v'
\,+\,D_Z \nabla^2 Z'.
\end{align}
Normal modes are then sought in the form
\begin{equation}
\bm{q}'(x,y,t)=\hat{\bm{q}}(y)\exp\!\left(i k_x x + \lambda t\right),
\label{eq:normal_mode}
\end{equation}
where $k_x$ is the streamwise wavenumber and $\lambda$ is the complex temporal eigenvalue. The real part $\Re(\lambda)$ is the temporal growth rate and the imaginary part determines the phase speed through
\begin{equation}
c_r = -\frac{\Im(\lambda)}{k_x}.
\end{equation}

Substitution of Eq.~\eqref{eq:normal_mode} into the linearized primitive-variable system produces a cross-stream eigenproblem of the form
\begin{equation}
\mathcal{L}(k_x)\hat{\bm{q}}=\lambda \hat{\bm{q}},
\label{eq:direct_problem}
\end{equation}
for the six-component state
\begin{equation}
\hat{\bm{q}}=
\left(
\hat{\rho},\,
\hat{u},\,
\hat{v},\,
\hat{w},\,
\hat{T},\,
\hat{Z}
\right)^{\mathsf{T}}.
\end{equation}
The operator retains the cross-stream variation of the reacting base state and includes simplified diffusive closures in the momentum, thermal, and scalar equations. Although the operator is evaluated numerically, its role is conceptually that of a finite-thickness temporal Kelvin--Helmholtz stability problem with variable thermochemical coefficients.

The eigenproblem is solved for a dense set of streamwise wavenumbers spanning the neighborhood of the imposed fundamental disturbance. Because several modes may coexist at a given $k_x$, the Kelvin--Helmholtz family used below is selected by anchoring the branch at the DNS fundamental wavenumber to the leading-growth mode at that wavenumber and then following the nearest continuation in combined eigenvalue/eigenvector space. This selected branch is the one used in all subsequent receptivity and modal projection results. A branch-selection sensitivity check is reported in Appendix~\ref{app:support_checks}.

\subsection{Adjoint problem and energy-weighted inner product}
The adjoint operator is defined through the weighted Lagrange identity
\begin{equation}
\langle \bm{q}^{\dagger},\mathcal{L}\bm{q}\rangle_E
-
\langle \mathcal{L}^{\dagger}\bm{q}^{\dagger},\bm{q}\rangle_E
=
\mathcal{B}(\bm{q}^{\dagger},\bm{q}),
\label{eq:lagrange_identity_paper}
\end{equation}
where $\mathcal{B}$ contains the outer-domain terms generated by integration by parts. We define the corresponding disturbance-energy-weighted inner product as
\begin{equation}
\langle \bm{a},\bm{b}\rangle_E
=\int \bm{a}^*(y)\,\mathsfbi{W}(y)\,\bm{b}(y)\,\mathrm{d}y,
\label{eq:energy_inner_product}
\end{equation}
with diagonal weight matrix
\begin{equation}
\mathsfbi{W}(y)=
\mathrm{diag}\!\left(
1,\,
\rho,\,
\rho,\,
\rho,\,
\rho c_v,\,
\rho c_v
\right).
\end{equation}
This choice weights the velocity components by density and the thermal/scalar components by $\rho c_v$, while leaving the density component unweighted. In the discrete calculation the quadrature-weighted metric is
\begin{equation}
\mathsfbi{M}=\mathrm{diag}\!\left(\mathsfbi{W}(y_i)\Delta y_i\right),
\qquad
\mathsfbi{L}^{\dagger}=\mathsfbi{M}^{-1}\mathsfbi{L}^{\mathsf{H}}\mathsfbi{M}.
\label{eq:discrete_metric_adjoint}
\end{equation}
The left eigenvectors of $\mathsfbi{L}$ are therefore transformed by $\mathsfbi{M}^{-1}$ before being used as energy-adjoint modes in the receptivity and projection calculations. It is closely aligned with the use of compressible disturbance-energy inner products descended from the formulation of Chu \citep{Chu1965} and later adjoint stability analyses of compressible flows \citep{Hill1995,Schmid2007}. For the temporal mixing layer, periodicity removes the streamwise boundary contributions, while the truncated outer-domain terms are controlled numerically by domain placement and sponge damping. The adjoint eigenproblem is therefore written as
\begin{equation}
\mathcal{L}^{\dagger}(k_x)\hat{\bm{q}}^{\dagger}
=\lambda^* \hat{\bm{q}}^{\dagger},
\label{eq:adjoint_problem}
\end{equation}
where the dagger denotes the adjoint under this weighted inner product.

For each $k_x$, direct and adjoint modes are paired by proximity of their eigenvalues under complex conjugation and then normalized so that
\begin{equation}
\langle \hat{\bm{q}}_j^{\dagger},\hat{\bm{q}}_j\rangle_E = 1.
\label{eq:biorth_normalization}
\end{equation}
The corresponding overlap and biorthogonality-error diagnostics are
\begin{equation}
\mathcal{O}_j(k_x)=
\left|\langle \hat{\bm{q}}_j^{\dagger},\hat{\bm{q}}_j\rangle_E\right|,
\qquad
\varepsilon_j(k_x)=
\left|\langle \hat{\bm{q}}_j^{\dagger},\hat{\bm{q}}_j\rangle_E-1\right|.
\end{equation}
The normalized direct and adjoint eigenpair residuals used in the verification plots are
\begin{equation}
\eta_j(k_x)=
\frac{\left\|\mathcal{L}(k_x)\hat{\bm{q}}_j-\lambda_j\hat{\bm{q}}_j\right\|_2}
{\left\|\mathcal{L}(k_x)\right\|_2\left\|\hat{\bm{q}}_j\right\|_2},
\end{equation}
\begin{equation}
\eta_j^\dagger(k_x)=
\frac{\left\|\mathcal{L}^\dagger(k_x)\hat{\bm{q}}_j^\dagger-\lambda_j^*\hat{\bm{q}}_j^\dagger\right\|_2}
{\left\|\mathcal{L}^\dagger(k_x)\right\|_2\left\|\hat{\bm{q}}_j^\dagger\right\|_2}.
\end{equation}
These quantities serve both as checks on the modal pairing and as summaries of how well the weighted direct--adjoint system behaves across the selected branch.

\section{Receptivity and Biorthogonal Decomposition Framework}
\subsection{Localized forcing channels}
Let the linearized temporal system be written as
\begin{equation}
\partial_t \bm{q}' = \mathcal{L}\bm{q}' + \bm{f},
\end{equation}
with forcing $\bm{f}(x,y,t)$. For a direct/adjoint mode pair $(\hat{\bm{q}}_j,\hat{\bm{q}}_j^\dagger)$ satisfying Eq.~\eqref{eq:biorth_normalization}, the corresponding modal forcing coefficient is
\begin{equation}
\mathcal{R}_j = \langle \hat{\bm{q}}_j^\dagger,\hat{\bm{f}}\rangle_E.
\label{eq:receptivity_projection}
\end{equation}
In the single-mode harmonic limit, the forced amplitude therefore scales as
\begin{equation}
a_j \propto
\frac{\langle \hat{\bm{q}}_j^\dagger,\hat{\bm{f}}\rangle_E}
{\lambda-\lambda_j}.
\end{equation}
For a localized forcing channel centered at a cross-stream location $y_f$, the receptivity map is consequently represented by the weighted magnitude of the adjoint component associated with that forcing variable.

The present paper retains five forcing families:
\begin{equation}
\bm{f}_{\rho},\qquad
\bm{f}_{u},\qquad
\bm{f}_{v},\qquad
\bm{f}_{T},\qquad
\bm{f}_{Z},
\end{equation}
which we refer to as mass, streamwise-momentum, transverse-momentum, thermal, and mixture-fraction forcing, respectively. In addition, the thermal and mixture-fraction maps are reported both in unweighted form and in chemistry-weighted form by multiplying the cross-stream maps by the magnitudes of the tabulated thermal and composition source proxies evaluated along the base state. These weighted variants are not new forcing operators in the strict linear sense; rather, they are physics-informed reweightings used to emphasize where the reacting base state is most thermochemically active.

\subsection{Biorthogonal modal amplitudes}
Given a direct/adjoint basis $\{\hat{\bm{q}}_j,\hat{\bm{q}}_j^{\dagger}\}$ that satisfies the normalization in Eq.~\eqref{eq:biorth_normalization}, the modal amplitude associated with a state vector $\bm{s}$ is
\begin{equation}
a_j = \langle \hat{\bm{q}}_j^{\dagger}, \bm{s}\rangle_E.
\label{eq:modal_amplitude}
\end{equation}
For a truncated basis containing the selected mode and the next $M-1$ retained modes, the biorthogonal reconstruction is
\begin{equation}
\bm{s}^{(M)}=\sum_{j=0}^{M-1} a_j \hat{\bm{q}}_j.
\label{eq:biorth_reconstruction}
\end{equation}

Because the direct basis is not orthogonal under the unweighted Euclidean metric, the amplitudes $a_j$ should not be interpreted as orthogonal energy fractions. Instead, the present work uses two more robust summaries. The first is the single-mode amplitude history $|a_j(t)|$ for the selected Kelvin--Helmholtz branch. The second is a cumulative energy-inner-product correlation between the data state and its truncated reconstruction,
\begin{equation}
\mathcal{C}^{(M)}(t)=
\frac{\left|\langle \bm{s}^{(M)},\bm{s}\rangle_E\right|}
{\sqrt{\langle \bm{s}^{(M)},\bm{s}^{(M)}\rangle_E
\langle \bm{s},\bm{s}\rangle_E}},
\label{eq:cumulative_correlation}
\end{equation}
which is bounded and therefore easier to interpret than a non-orthogonal energy-partition measure.

\subsection{DNS planar modal projection}
The biorthogonal decomposition is applied to time-resolved planar DNS data at the streamwise fundamental of the selected branch. For each saved slice, the primitive perturbation fields
\begin{equation}
\rho',\quad u',\quad v',\quad w',\quad T',\quad Z'
\end{equation}
are formed by subtracting the mean base profiles. The streamwise Fourier coefficient at the selected wavenumber $k_x$ is then computed as
\begin{equation}
\hat{f}(y,t;k_x)=
\frac{1}{L_x}\int_0^{L_x} f'(x,y,t)\,e^{-ik_x x}\,\mathrm{d}x.
\label{eq:fourier_projection}
\end{equation}
Stacking these six Fourier coefficients produces a time-dependent state vector $\bm{s}(t;k_x)$ defined on the cross-stream grid of the modal operator.

All reported histories are nondimensionalized using the characteristic temporal scale
\begin{equation}
t^* = t\,\frac{U_{c,0}}{\delta_{\omega,0}},
\end{equation}
where
\begin{equation}
U_{c,0}=\frac{U_t+U_b}{2}
\end{equation}
is the arithmetic-mean outer-stream velocity of the reference base state and $\delta_{\omega,0}\equiv \delta_{\omega,\mathrm{ref}}$. This scaling allows the modal-projection time histories to be interpreted against the characteristic turnover time of the mean reacting shear layer.

\section{Theory Reference}
\subsection{Compressible vortex-sheet limit}
To interpret the finite-thickness numerical branch, we also consider a compressible vortex-sheet reference constructed from the outer-stream states of the reacting temporal mixing layer. The upper and lower streams are taken to be uniform with states
\begin{equation}
(\rho_t,U_t,a_t),\qquad (\rho_b,U_b,a_b),
\end{equation}
where the density, velocity, and sound speed are extracted by averaging the outer portions of the numerical base state. The temporal normal-mode ansatz is written as
\begin{equation}
q' \sim e^{i(k_x x-\omega t)},
\end{equation}
with evanescent cross-stream dependence in each stream. The corresponding decay rates are
\begin{equation}
\alpha_j^2 = k_x^2 - \frac{(\omega-k_x U_j)^2}{a_j^2},
\qquad j\in\{t,b\}.
\end{equation}
Enforcing the interface pressure and kinematic conditions yields the classical compressible vortex-sheet dispersion relation
\begin{equation}
\frac{\rho_t(\omega-k_x U_t)^2}{\alpha_t}
\;+\;
\frac{\rho_b(\omega-k_x U_b)^2}{\alpha_b}
=0.
\label{eq:vortex_sheet_dispersion}
\end{equation}
The roots of Eq.~\eqref{eq:vortex_sheet_dispersion} provide the reference temporal frequency $\omega(k_x)$ from which phase speed and growth rate are extracted. For the convective Mach numbers quoted later, we use the density-weighted convective speed
\begin{equation}
U_{\mathrm{conv}}=
\frac{\rho_t U_t+\rho_b U_b}{\rho_t+\rho_b},
\qquad
M_{c,j}=\frac{|U_j-U_{\mathrm{conv}}|}{a_j},
\qquad j\in\{t,b\}.
\end{equation}
This velocity is used only to characterize the outer-stream compressible reference state. It is distinct from the arithmetic-mean velocity $U_{c,0}$ used to define the temporal scale $t^*$.

\subsection{Connection to the finite-thickness numerical branch}
The numerical eigenproblem uses the convention $\exp(i k_x x + \lambda t)$, whereas the vortex-sheet reference uses $\exp(i k_x x - \omega t)$. The two are related by
\begin{equation}
\lambda = -i\omega,
\end{equation}
so that
\begin{equation}
\sigma = \Re(\lambda) = \Im(\omega),
\qquad
c_r = -\frac{\Im(\lambda)}{k_x} = \frac{\Re(\omega)}{k_x}.
\end{equation}
The vortex-sheet model is used as an idealized outer-state reference. Its role is to indicate how strongly the finite-thickness numerical branch departs from the discontinuous compressible limit built from the same upper and lower streams.

\section{Results}
\subsection{Base-state and thermo-transport structure}
The mean reacting base state is organized by strong but smooth cross-layer variation in all of the primary primitive variables. The lower stream is dense, cool, and fuel-rich, with $\bar{\rho}_b^*=1$, $\tilde{u}_b^*=-0.252$, $\tilde{T}_b^*=1$, and $\tilde{Z}_b\approx 1$, whereas the upper stream is lighter, faster, hotter, and oxidizer-rich, with $\bar{\rho}_t^*=0.324$, $\tilde{u}_t^*=0.748$, $\tilde{T}_t^*=3.023$, and $\tilde{Z}_t\approx 0$. Because these changes occur continuously over the finite reference vorticity thickness, the stability problem is governed by a reacting variable-coefficient shear profile with no reduction to a discontinuous or constant-property approximation.

\begin{figure}
  \centering
  \includegraphics[width=0.78\textwidth]{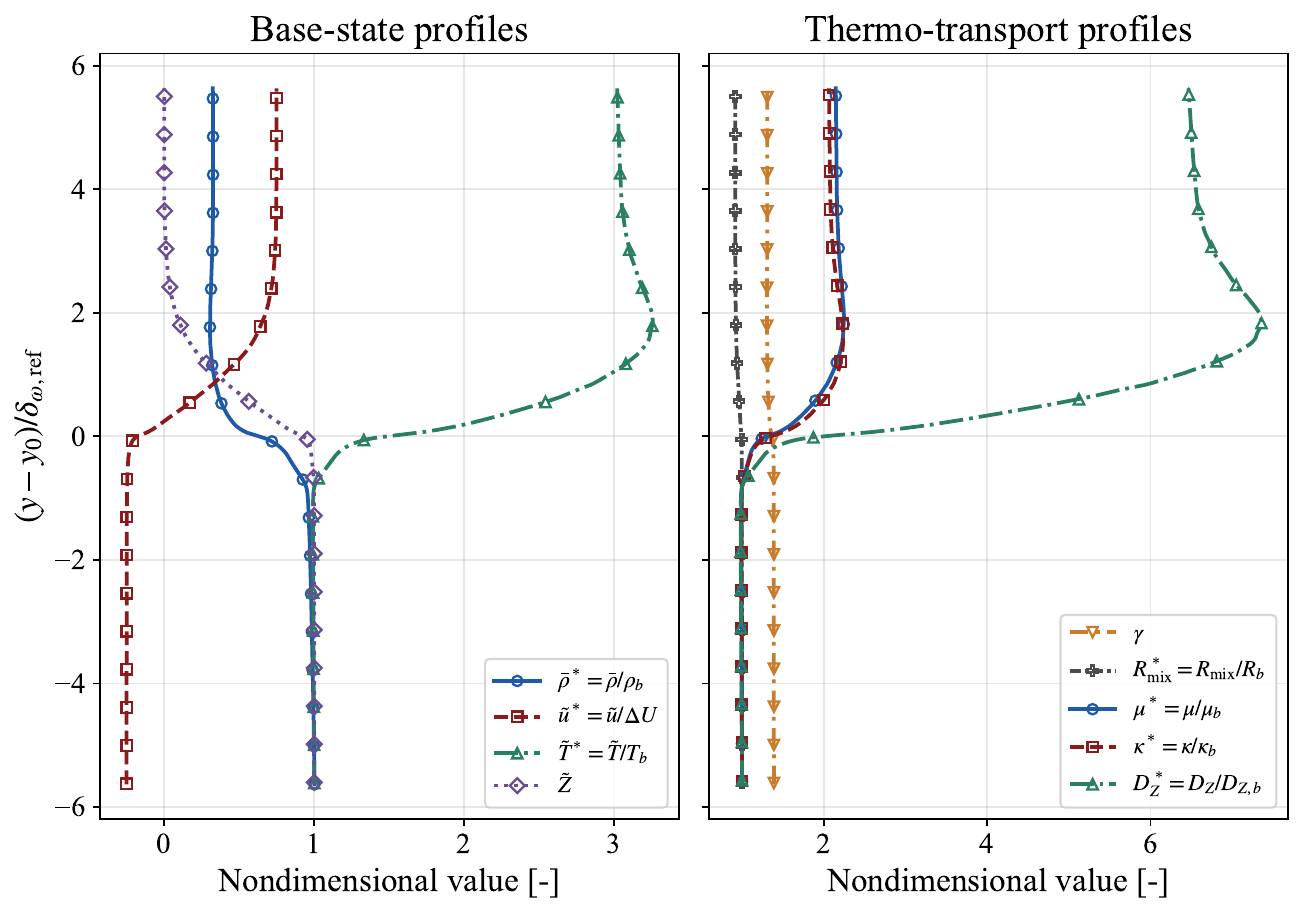}
  \caption{Scaled base-state and thermo-transport profiles. The first panel summarizes the mean reacting layer through $\bar{\rho}^*$, $\tilde{u}^*$, $\tilde{T}^*$, and $\tilde{Z}$, while the second panel collects the thermodynamic and transport quantities that close the variable-coefficient temporal operator.}
  \label{fig:base_profiles}
\end{figure}

The thermodynamic and transport closures inherit the same smooth but strongly inhomogeneous cross-stream structure. Across the layer, $\gamma$ varies from $1.306$ to $1.391$, $R_{\mathrm{mix}}^*$ from $0.917$ to $1.000$, $\mu^*$ from $0.993$ to $2.248$, $\kappa^*$ from $0.992$ to $2.238$, and $D_Z^*$ from $0.982$ to $7.353$. These profiles set the local compressibility, diffusive, and thermochemical couplings that enter the temporal operator. The nondimensional presentation in Fig.~\ref{fig:base_profiles} is therefore a compact view of the coefficient field that filters every direct mode, adjoint mode, receptivity map, and modal projection considered below.

\subsection{Selected KH branch and direct/adjoint mode structure}
The eigenvalue sweep over $k_x$ produces a broad cloud of temporal modes, from which the selected Kelvin--Helmholtz family is identified by anchoring at the DNS fundamental and tracking outward in wavenumber. This family is unstable over two low-to-moderate nondimensional wavenumber intervals, $0.453\lesssim k_x^*\lesssim 0.511$ and $0.627\lesssim k_x^*\lesssim 0.801$. Its maximum nondimensional growth rate is $\sigma^*\approx 7.35\times 10^{-3}$ at $k_x^*\approx 0.627$, with associated nondimensional phase speed $c_r^*\approx 1.56\times 10^{-2}$. At the wavenumber closest to the imposed fundamental, $k_x^*\approx 0.569$, the selected branch is slightly damped, with $\sigma^*\approx -9.47\times 10^{-4}$ and $c_r^*\approx 1.80\times 10^{-2}$. Over the full selected set, the nondimensional phase speed spans approximately $-2.39\times 10^{-2}$ to $1.99\times 10^{-2}$, showing that the reacting finite-thickness branch propagates much more slowly than the outer-stream velocity difference and in places changes sign relative to the mean-shear scaling. These trends indicate that the instability is controlled by the distributed shear layer and cannot be reduced to simple advection with either outer state. In broad terms, this is consistent with the compressible-mixing-layer literature, in which increasing compressibility suppresses growth without eliminating the role of a dominant Kelvin--Helmholtz family
\citep{PapamoschouRoshko1988,SandhamReynolds1990,SandhamReynolds1991,VremanSandhamLuo1996,FreundLeleMoin2000,FoysiSarkar2010}. It is
also consistent with inviscid compressible stability calculations in which the location, propagation speed, and multiplicity of unstable branches depend sensitively on the thermodynamic structure used to model the layer
\citep{Blumen1970,BlumenDrazinBillings1975,DrazinDavey1977,JacksonGrosch1989}.

The direct and adjoint mode shapes at the fundamental wavenumber provide the cross-stream structure needed to interpret that branch physically. The direct mode describes how the unstable disturbance is distributed among density, velocity, temperature, and mixture-fraction perturbations, whereas the adjoint mode identifies the regions in which forcing couples most efficiently into the same instability. Read together, the two mode families show that the reacting Kelvin--Helmholtz response reflects coupled kinematic and thermochemical support. It is localized by the finite-thickness shear layer, and the modal support also samples the same temperature and composition gradients that shape the transport and thermodynamic coefficients in Fig.~\ref{fig:base_profiles}. The selected branch therefore furnishes the modal backbone for both the receptivity maps and the later DNS projections.
\begin{figure}
  \centering
  \begin{subfigure}{0.48\textwidth}
    \centering
    \includegraphics[width=\textwidth]{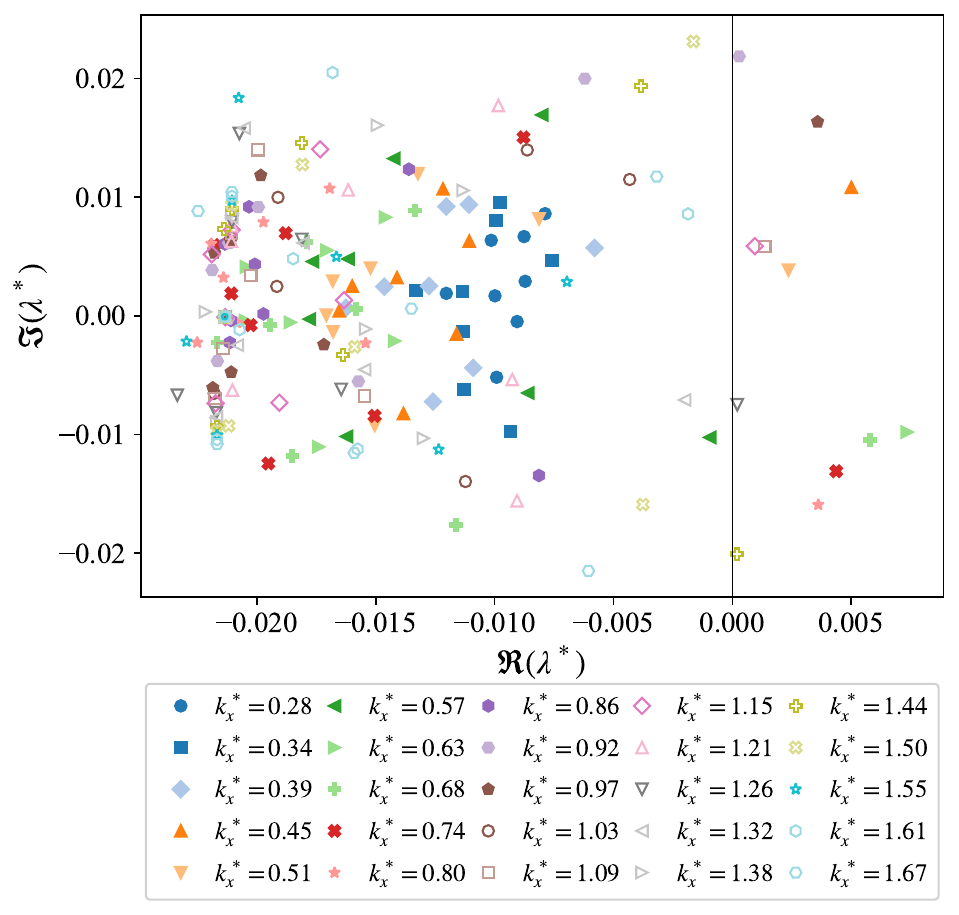}
    \caption{Full temporal eigenspectrum over the $k_x$ sweep.}
  \end{subfigure}\hfill
  \begin{subfigure}{0.48\textwidth}
    \centering
    \includegraphics[width=\textwidth]{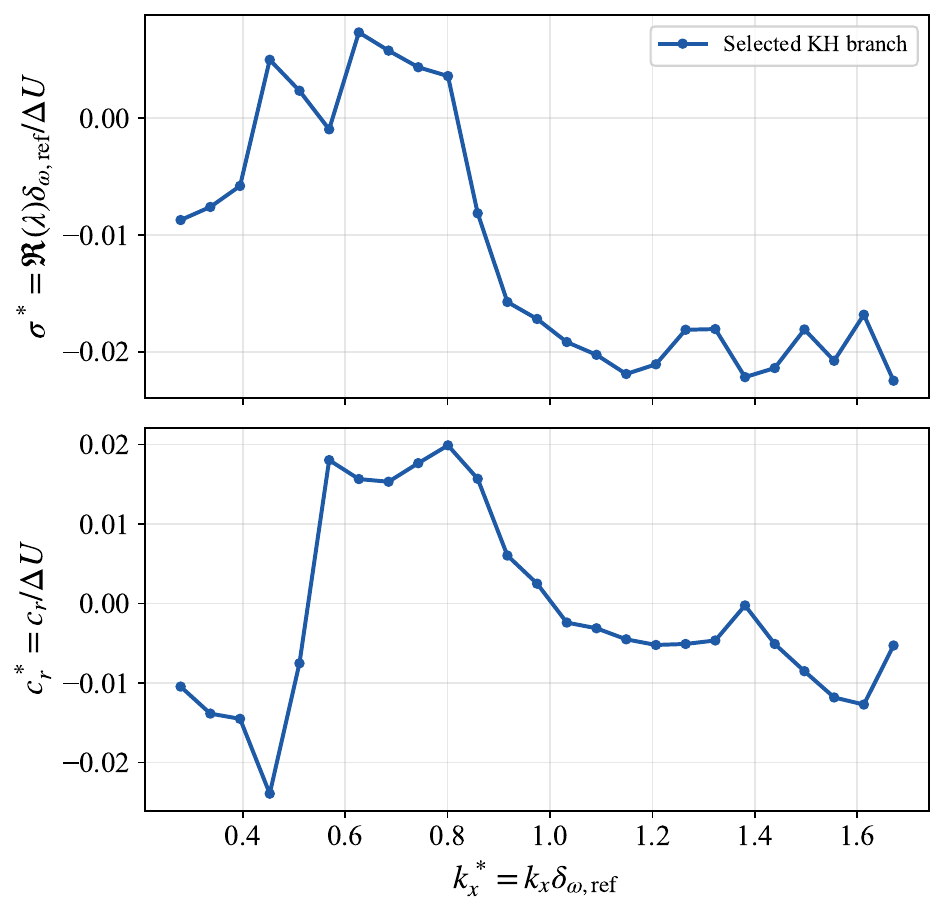}
    \caption{Selected Kelvin--Helmholtz branch dispersion.}
  \end{subfigure}
  \caption{Spectrum-level view of the temporal instability problem. The full eigenspectrum shows the broader numerical family, while the selected branch isolates the mode family used throughout the receptivity analysis.}
  \label{fig:spectrum_branch}
\end{figure}

\begin{figure}
  \centering
  \begin{subfigure}{0.48\textwidth}
    \centering
    \includegraphics[width=\textwidth]{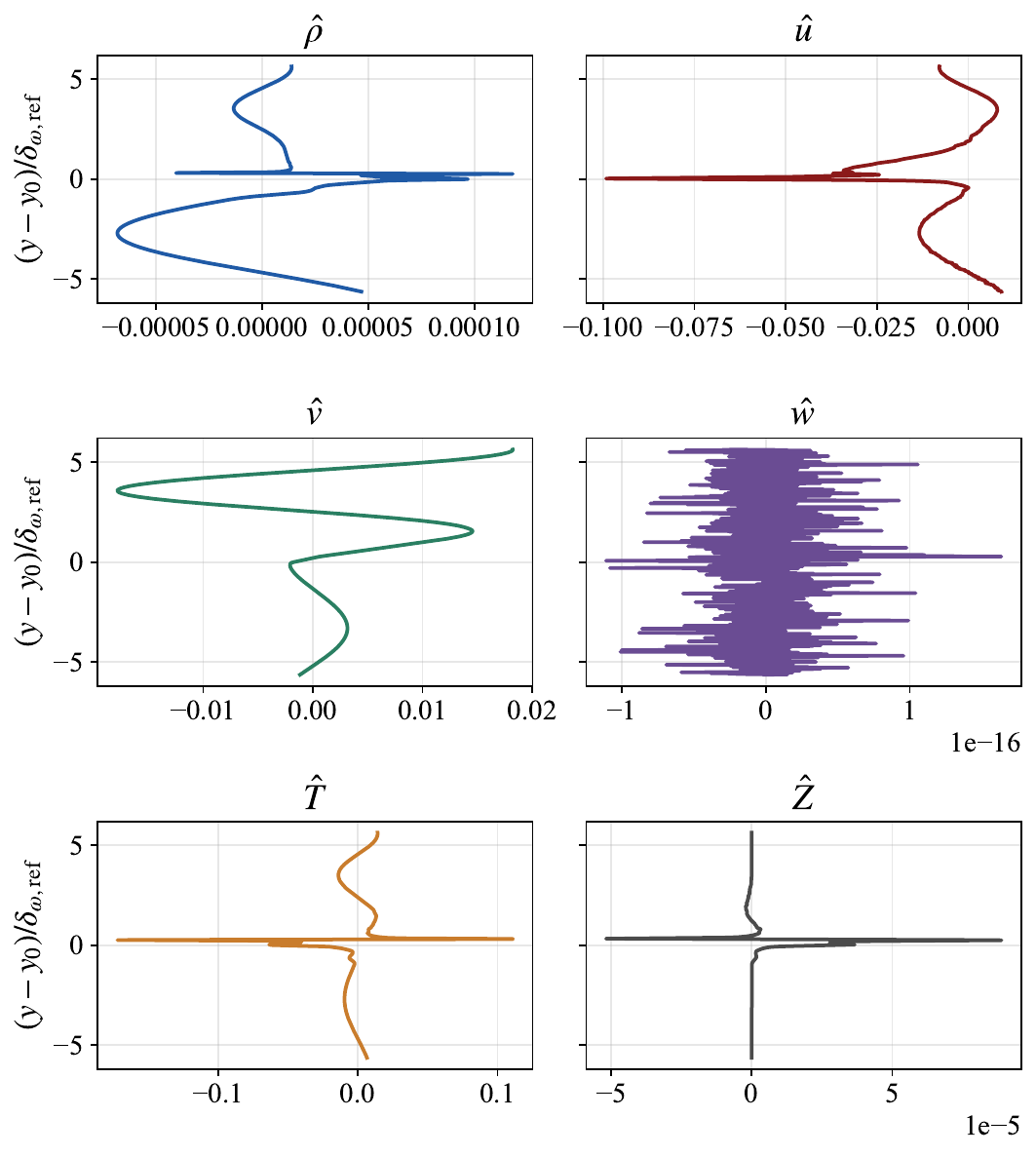}
    \caption{Direct mode shape of the selected branch.}
  \end{subfigure}\hfill
  \begin{subfigure}{0.48\textwidth}
    \centering
    \includegraphics[width=\textwidth]{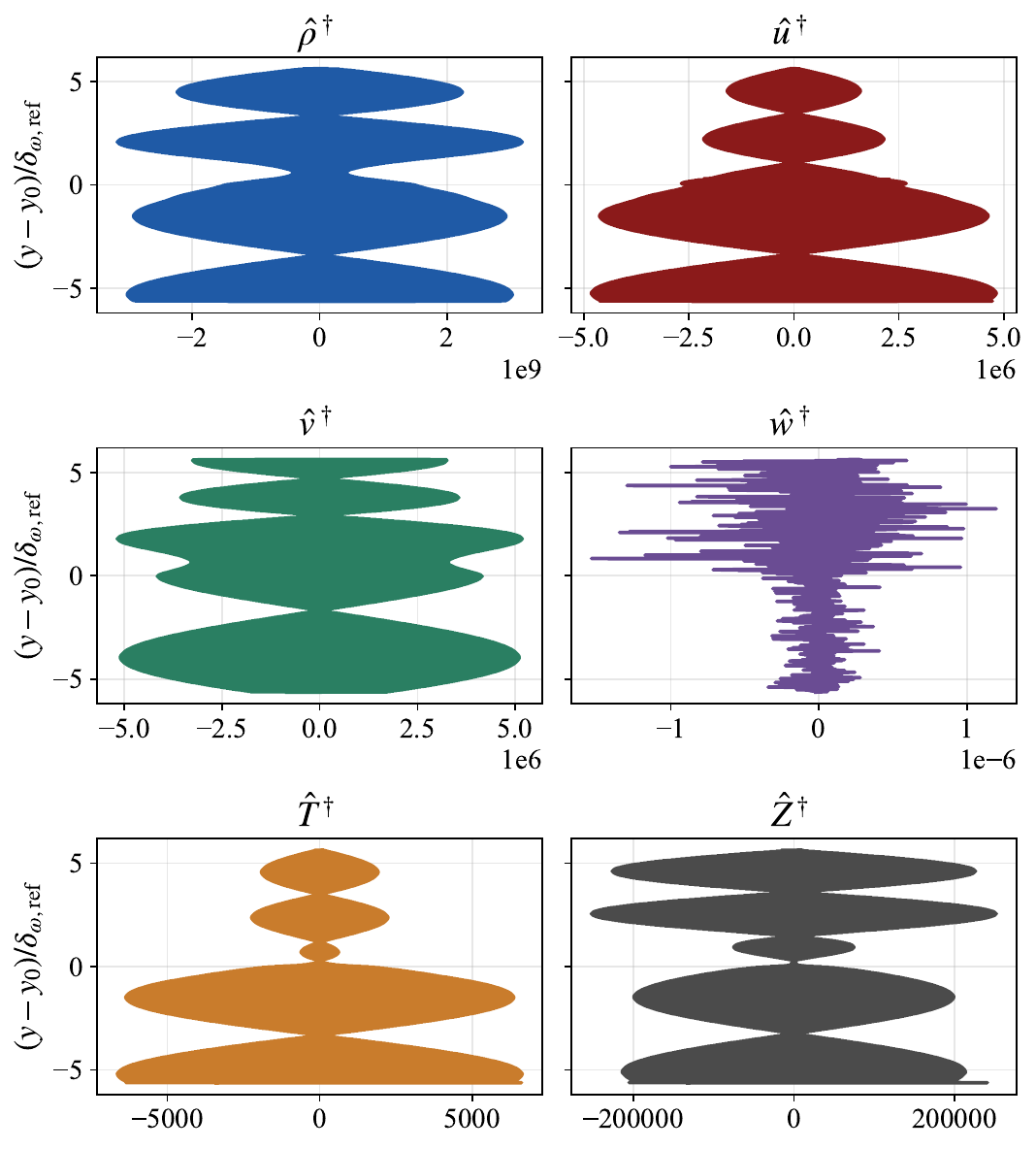}
    \caption{Adjoint mode shape of the selected branch.}
  \end{subfigure}
  \caption{Direct and adjoint cross-stream structures at the fundamental streamwise wavenumber. The direct mode describes the response shape of the selected instability branch, while the adjoint mode indicates the cross-stream regions in which localized forcing most efficiently excites it.}
  \label{fig:mode_shapes}
\end{figure}

\subsection{Biorthogonality and receptivity maps}
\begin{figure}
  \centering
  \begin{subfigure}{0.32\textwidth}
    \centering
    \includegraphics[width=\textwidth]{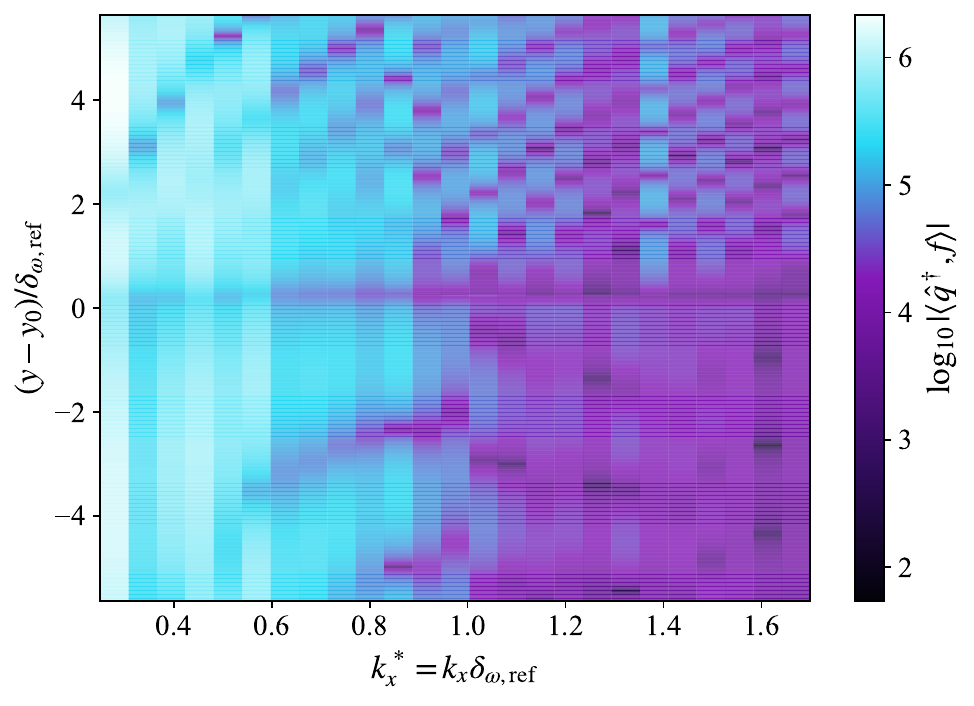}
    \caption{Overall Mass forcing.~\\~}
  \end{subfigure}\hfill
  \begin{subfigure}{0.32\textwidth}
    \centering
    \includegraphics[width=\textwidth]{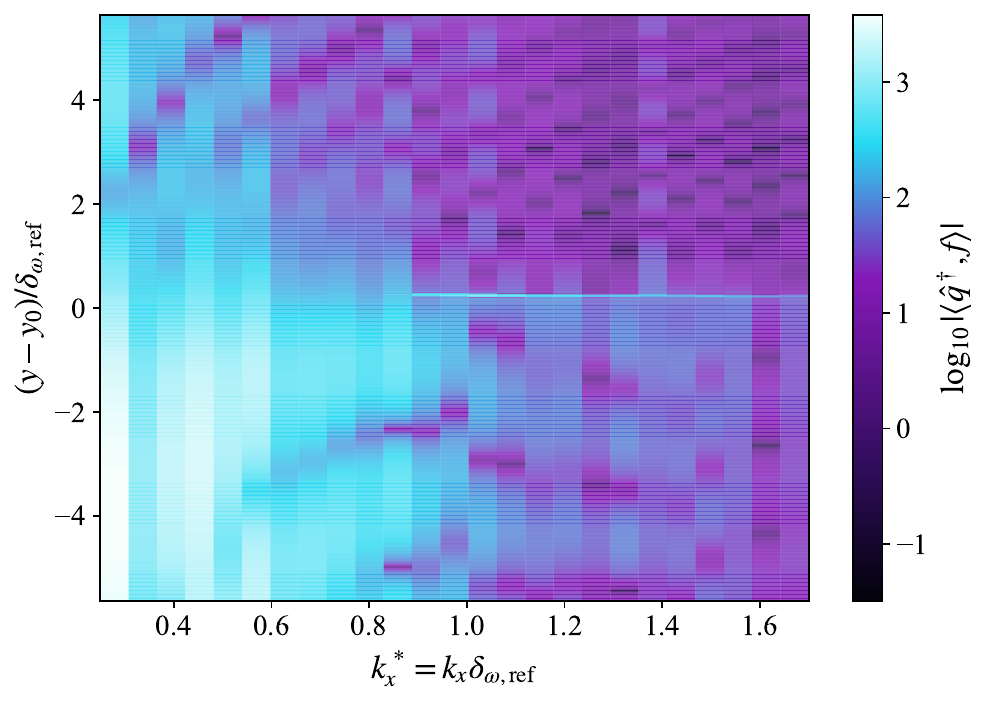}
    \caption{Streamwise momentum forcing.}
  \end{subfigure}\hfill
  \begin{subfigure}{0.32\textwidth}
    \centering
    \includegraphics[width=\textwidth]{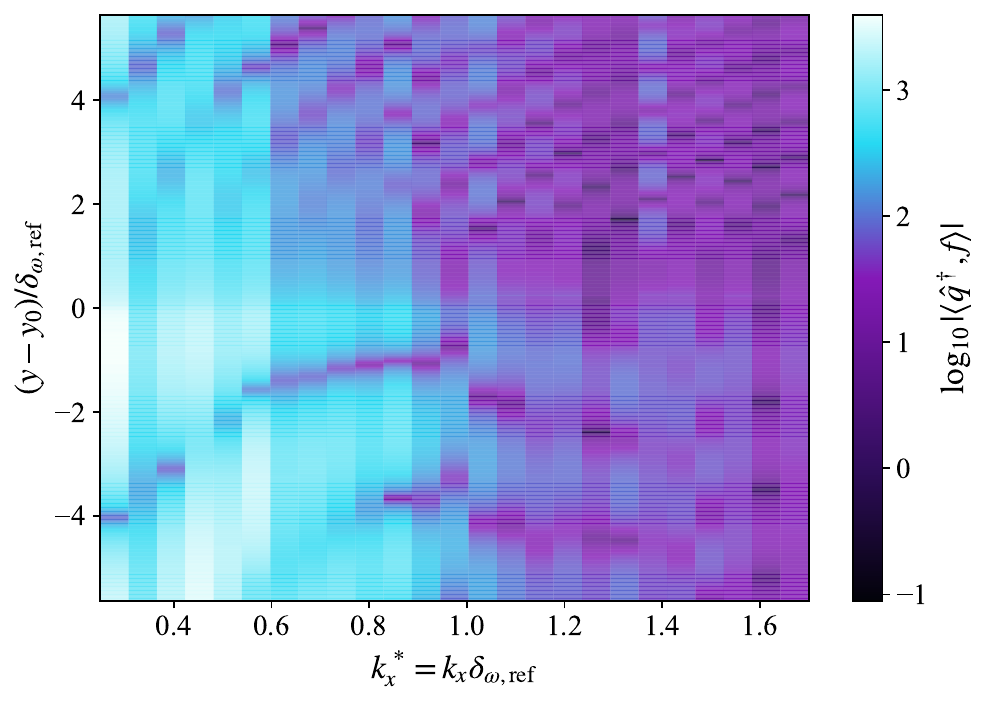}
    \caption{Transverse momentum forcing.}
  \end{subfigure}\\[0.5em]
  \begin{subfigure}{0.32\textwidth}
    \centering
    \includegraphics[width=\textwidth]{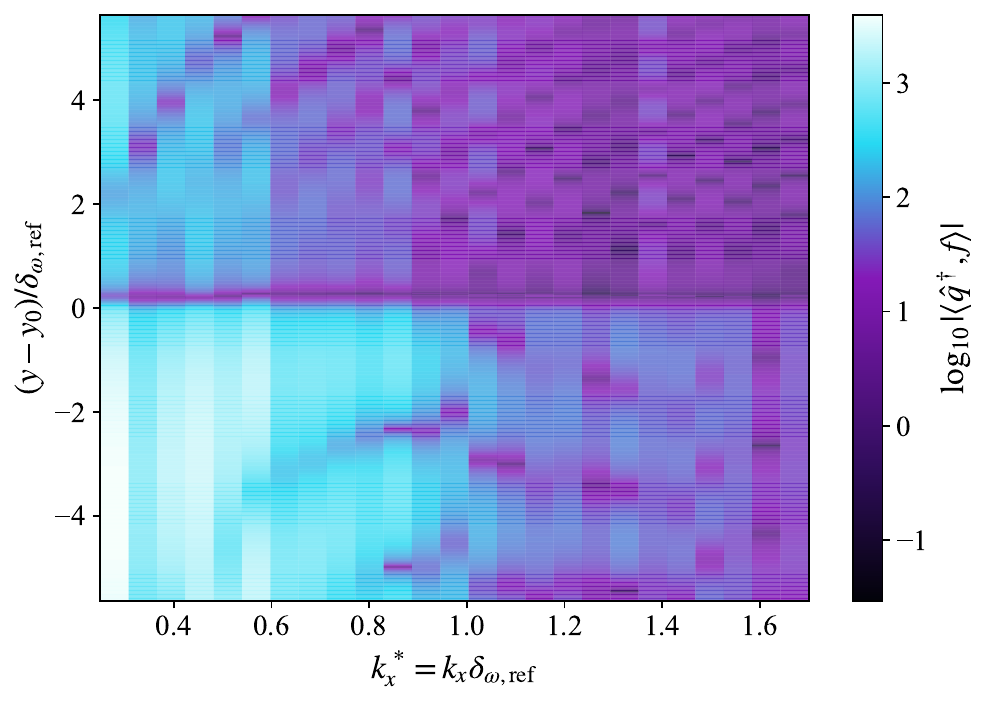}
    \caption{Overall Thermal forcing.~\\~}
  \end{subfigure}\hfill
  \begin{subfigure}{0.32\textwidth}
    \centering
    \includegraphics[width=\textwidth]{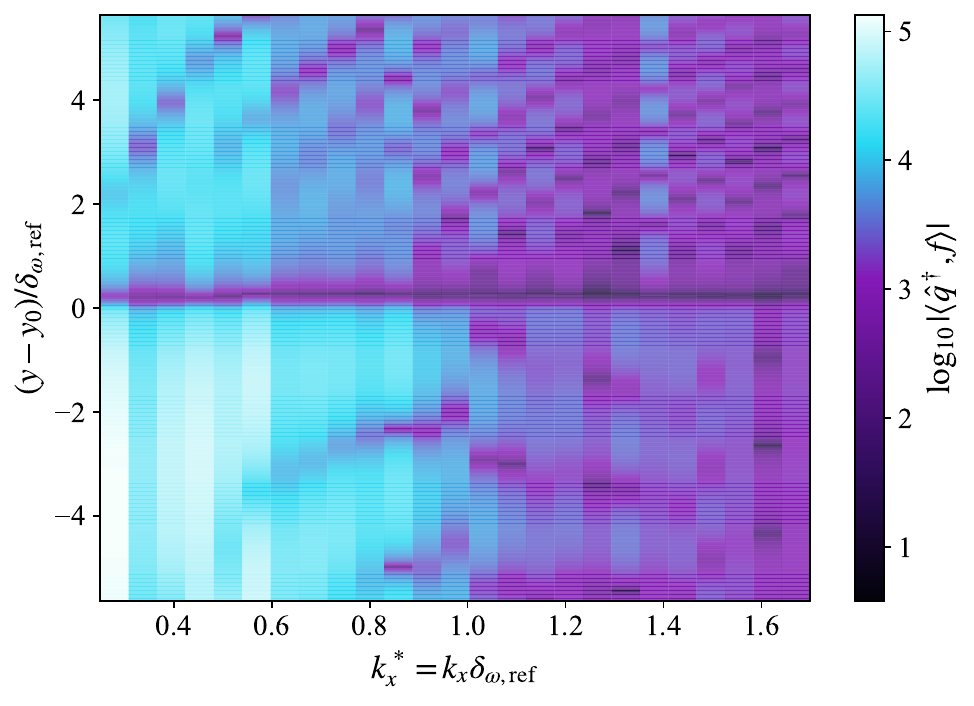}
    \caption{Overall $Z$ forcing.~\\~}
  \end{subfigure}\hfill
  \begin{subfigure}{0.32\textwidth}
    \centering
    \includegraphics[width=\textwidth]{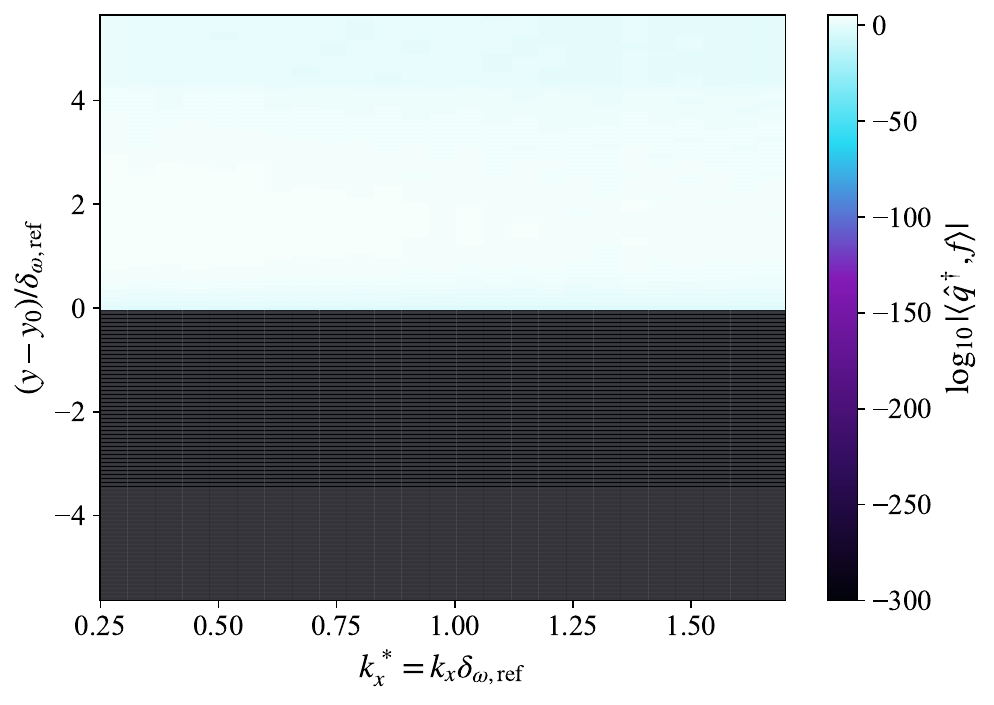}
    \caption{Chemistry-weighted thermal forcing.}
  \end{subfigure}
  \caption{Log-scale receptivity maps for the selected branch. The first five panels show the classical and reacting forcing channels directly from the adjoint projection, while the sixth panel illustrates a chemistry-weighted thermal variant.}
  \label{fig:receptivity_maps}
\end{figure}

The direct and adjoint branches remain well paired under the energy-weighted inner product. Over the full selected branch, the overlap is held at its target normalization and the biorthogonality error $\varepsilon_j$ remains between \num{8.06e-10} and \num{8.55e-06}. These values are small on the scale of the forcing projections, confirming that the branch pairing and normalization are numerically consistent over the entire $k_x$ range used for the receptivity analysis. This conditioning matters directly, because every forcing map reported in Fig.~\ref{fig:receptivity_maps} is constructed from the weighted adjoint projection and would otherwise inherit any pairing error. An independent residual check leads to the same conclusion. Over the first three retained modes, the normalized direct eigenpair residuals remain between
\num{3.35e-19} and \num{2.79e-16}, while the corresponding adjoint residuals
remain between \num{3.57e-19} and \num{3.92e-16}. The computed branch satisfies both the overlap normalization and the direct and adjoint eigenproblems to high accuracy across the full $k_x$ sweep.

The receptivity maps show that the selected Kelvin--Helmholtz branch is most sensitive in the low-to-moderate wavenumber range and that the leading channel depends strongly on whether the quantity is viewed as a raw localized forcing kernel or as a thermochemical-support-weighted response. Using the cross-stream averaged maxima as a compact summary, the raw mass map is the largest unweighted channel, with peak magnitude $\approx 1.26\times 10^6$ at $k_x^*\approx 0.278$. The raw mixture-fraction map forms the second-largest unweighted channel, with peak magnitude $\approx 6.51\times 10^4$ at the same wavenumber, because composition perturbations enter the pressure closure through $p_Z Z'$. The streamwise-momentum, transverse-momentum, and thermal maps peak at $\approx 1.73\times 10^3$, $1.66\times 10^3$, and $1.72\times 10^3$, respectively. The global maxima give the same leading ordering: the mass map reaches $2.16\times 10^6$ near $y^*\approx 4.14$, whereas the largest unweighted mixture-fraction response reaches $1.34\times 10^5$ near $y^*\approx -3.93$.

The chemistry-weighted maps answer a different physical question by combining modal sensitivity with the thermochemical support of the mean reacting layer. With this weighting, the thermal map attains a cross-stream averaged maximum $\approx 1.16\times 10^4$ and a global maximum $1.36\times 10^5$ near $y^*\approx 1.48$, showing that thermal forcing remains the dominant thermochemical pathway where heat-release support is included. The weighted mixture-fraction map is smaller, with cross-stream averaged maximum $\approx 5.15\times 10^2$ near $k_x^*\approx 0.453$ and global maximum $5.97\times 10^3$ near $y^*\approx 1.53$. Taken together, these maps indicate that the reacting extension of the receptivity problem separates two effects: mass and mixture-fraction pressure coupling provide the strongest raw localized projections onto the selected branch, while heat release marks the strongest thermochemically active support for that branch. This is compatible with receptivity formulations that emphasize localized energy deposition as an effective excitation pathway in compressible mixing layers
\citep{Tam1978,Balsa1988,BaroneLele2005,WuTumin2019}, while extending that
viewpoint here to a finite-thickness reacting operator built from the mean reacting state. Physically, the result also aligns with earlier observations that heat release can reorganize reacting shear-layer structure, entrainment, and turbulent spectral content
\citep{HermansonDimotakis1989,McmurtryRileyMetcalfe1989,PantanoSarkarWilliams2003,KnausPantano2009,JahanbakhshiMadnia2018}. The present analysis resolves
that influence through a modal receptivity hierarchy and connects it directly to the instability branch rather than only to bulk growth or flame-shape measurements. The same receptivity formula can also be checked directly against a solved single-mode forced response. For a detuned forcing shift $\lambda_f$, the predicted and resolved single-mode amplitudes are
\begin{equation}
a_{0,\mathrm{pred}}
=
\frac{\langle \hat{\bm{q}}_0^\dagger,\bm{f}\rangle_E}{\lambda_f-\lambda_0},
\qquad
a_{0,\mathrm{solve}}
=
\left\langle \hat{\bm{q}}_0^\dagger,
(\lambda_f \mathsfbi{I}-\mathsfbi{L})^{-1}\bm{f}\right\rangle_E,
\end{equation}
and the verification quantity reported below is the absolute defect
\begin{equation}
\delta_f = \left|a_{0,\mathrm{solve}}-a_{0,\mathrm{pred}}\right|.
\end{equation}
Using a representative detuned harmonic solve at the fundamental wavenumber $k_x^*\approx 0.569$, Table~\ref{tab:forced_response} compares the predicted single-mode amplitude from the resolvent expression with the amplitude recovered from the resolved forced state. The agreement is excellent for all five forcing channels, with absolute defects between
\num{2.97e-09} and \num{1.85e-06}. The $Z$ row is now finite because the
composition-pressure term couples mixture-fraction forcing into the momentum equations.
\begin{figure}
  \centering
  \begin{subfigure}[t]{0.6\textwidth}
    \centering
    \includegraphics[width=\textwidth]{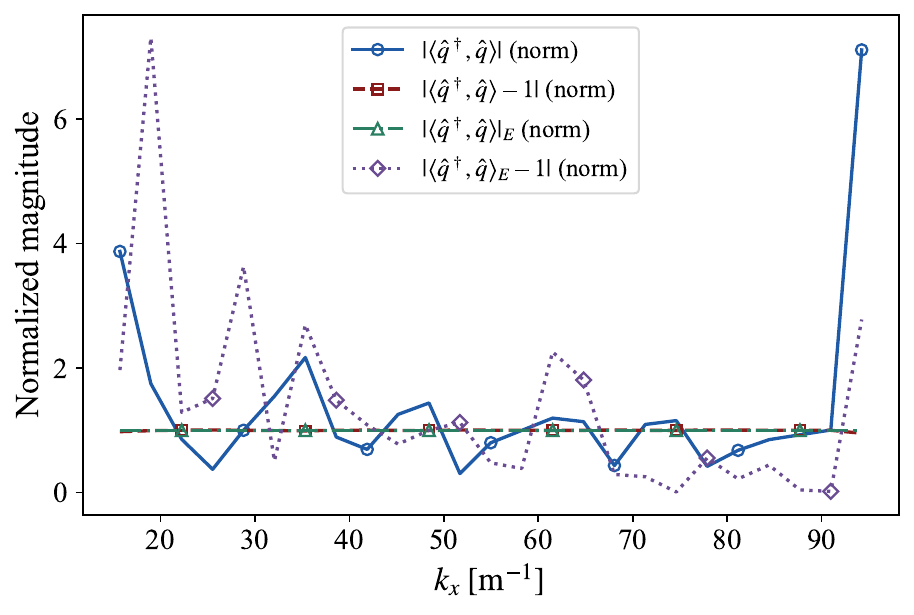}
    \caption{Energy-weighted overlap and biorthogonality error.}
  \end{subfigure}\hfill
  \begin{subfigure}[t]{0.62\textwidth}
    \centering
    \includegraphics[width=\textwidth]{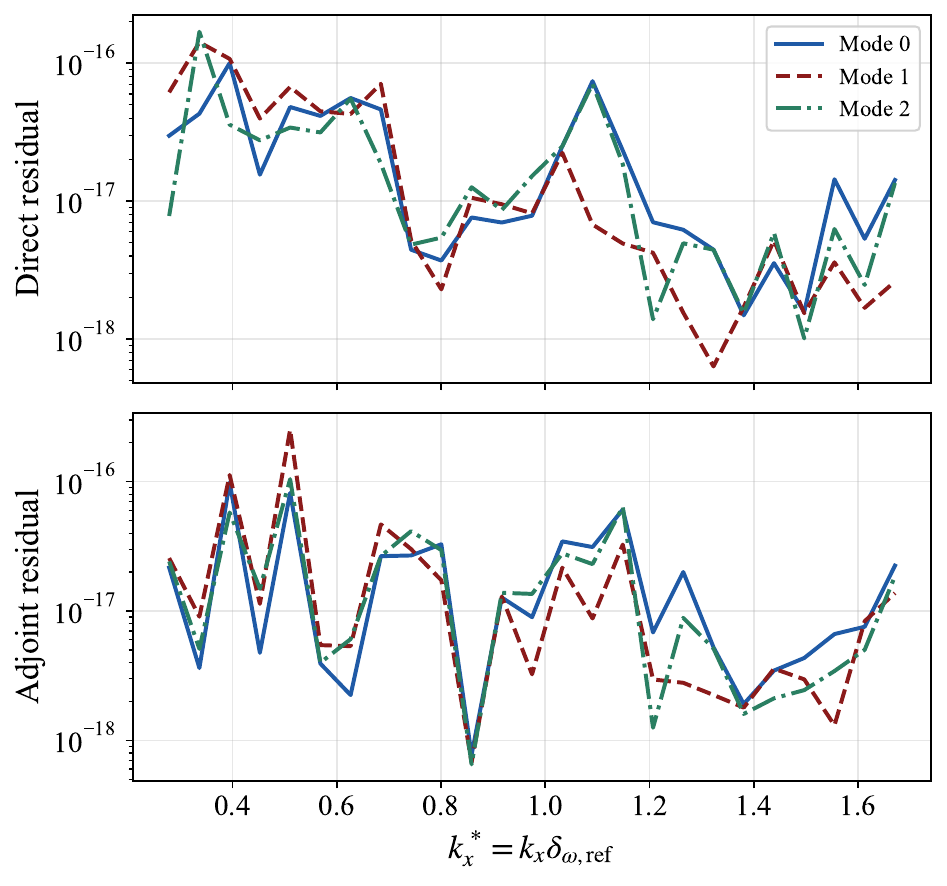}
    \caption{Direct and adjoint eigenpair residuals.}
  \end{subfigure}
  \caption{Verification diagnostics for the selected Kelvin--Helmholtz branch. The weighted overlap remains at the target normalization, and the direct and adjoint eigenpair residuals remain close to numerical precision over the full $k_x$ sweep.}
  \label{fig:biorth}
\end{figure}

\begin{table}[tbp]
  \centering
  \captionsetup{skip=2pt}
  \caption{Representative single-mode forced-response check at the fundamental wavenumber $k_x^*\approx 0.569$. The forcing location $y_f^*$ is chosen from the largest metric-consistent forcing kernel associated with each forcing channel.}
  \label{tab:forced_response}
  \begingroup
  \setlength{\tabcolsep}{7pt}
  \renewcommand{\arraystretch}{1.18}
  \rowcolors{2}{tablerow}{white}
  \begin{tabular}{@{}l
                  S[table-format=-1.3]
                  S[table-format=3.8e1]
                  S[table-format=3.8e1]
                  S[table-format=1.8e1]@{}}
    \toprule
    \rowcolor{tablehead}
    \multicolumn{1}{@{}l}{\textbf{Channel}} &
    \multicolumn{1}{c}{\textbf{$y_f^*$}} &
    \multicolumn{1}{c}{\textbf{$|a_0|_{\mathrm{pred}}$}} &
    \multicolumn{1}{c}{\textbf{$|a_0|_{\mathrm{solve}}$}} &
    \multicolumn{1}{c@{}}{\textbf{$|a_0^{\mathrm{solve}}-a_0^{\mathrm{pred}}|$}} \\
    \midrule
    $\rho$ & 2.517 & \num{3.81570218e2} & \num{3.81570217e2} & \num{1.85377440e-6} \\
    $u$ & -1.978 & \num{6.10066706e-1} & \num{6.10066703e-1} & \num{2.96537491e-9} \\
    $v$ & -4.181 & \num{7.98011522e-1} & \num{7.98011519e-1} & \num{3.88474620e-9} \\
    $T$ & -1.978 & \num{6.75005874e-1} & \num{6.75005872e-1} & \num{3.28241351e-9} \\
    $Z$ & -2.066 & \num{2.75533100e1} & \num{2.75533099e1} & \num{1.33932535e-7} \\
    \addlinespace[2pt]
    \bottomrule
  \end{tabular}
  \endgroup
\end{table}

\subsection{Time-resolved modal amplitudes and few-mode decomposition}
The planar DNS projection places the receptivity framework in direct contact with the time-resolved nonlinear data. The selected mode-$0$ amplitude remains finite throughout the sampled interval $0\le t^*\le 47.87$, with minimum, mean, and maximum values of approximately $117$, $2.43\times 10^3$, and $1.90\times 10^4$, respectively. The corresponding mode--state correlation $\mathcal{C}^{(1)}$ ranges from $0.117$ to $0.526$, with mean $0.375$. The selected Kelvin--Helmholtz branch therefore captures a persistent but incomplete portion of the fundamental streamwise content of the nonlinear planar data. It is clearly present, but it does not constitute a nearly closed one-mode description of the projected state. A spanwise representativeness check using the post-developing three-dimensional snapshots is reported in Appendix~\ref{app:support_checks}.

The componentwise contributions show that the projected modal amplitude is shared primarily between streamwise velocity, density, temperature, and cross-stream velocity. Averaged over the time series, the $\hat{u}$ contribution is $\approx 2.04\times 10^3$, followed by $\hat{\rho}\approx 6.25\times 10^2$, $\hat{T}\approx 5.98\times 10^2$, and $\hat{v}\approx 4.24\times 10^2$; the $\hat{w}$ and $\hat{Z}$ contributions are small at the selected planar fundamental. This ranking is consistent with the corrected receptivity maps: the raw projection is strongly kinematic and density weighted, while thermal content remains important where the branch samples the reacting thermodynamic structure. The few-mode amplitudes show additional nontrivial content in other retained modes, with mean amplitudes of approximately $2.42\times 10^3$, $3.73\times 10^3$, and $1.11\times 10^3$ for modes $1$, $2$, and $3$, respectively. However, the cumulative correlation based on the truncated biorthogonal reconstruction does not increase monotonically with retained mode count, remaining near $0.375$, $0.220$, $0.166$, and $0.166$ for $M=1,2,3,4$. This behavior reflects the non-orthogonality of the finite-dimensional direct basis and reinforces the need to interpret the reconstruction in biorthogonal, not orthogonal energy-partition, terms. The use of biorthogonal projections as a diagnostic bridge between DNS data and linear receptivity theory follows the same logic used in direct/adjoint modal interpretations of high-speed disturbance fields
\citep{TuminWangZhong2007,AlHasnineRussoTuminBrehm2023} and is consistent with
the broader modal-analysis view that linear coherent structures can remain detectable in complex nonlinear data without furnishing a complete reduced description \citep{Schmid2007,TairaEtAl2017}.
\begin{figure}
  \centering
  \begin{subfigure}[t]{0.42\textwidth}
    \centering
    \includegraphics[width=\textwidth]{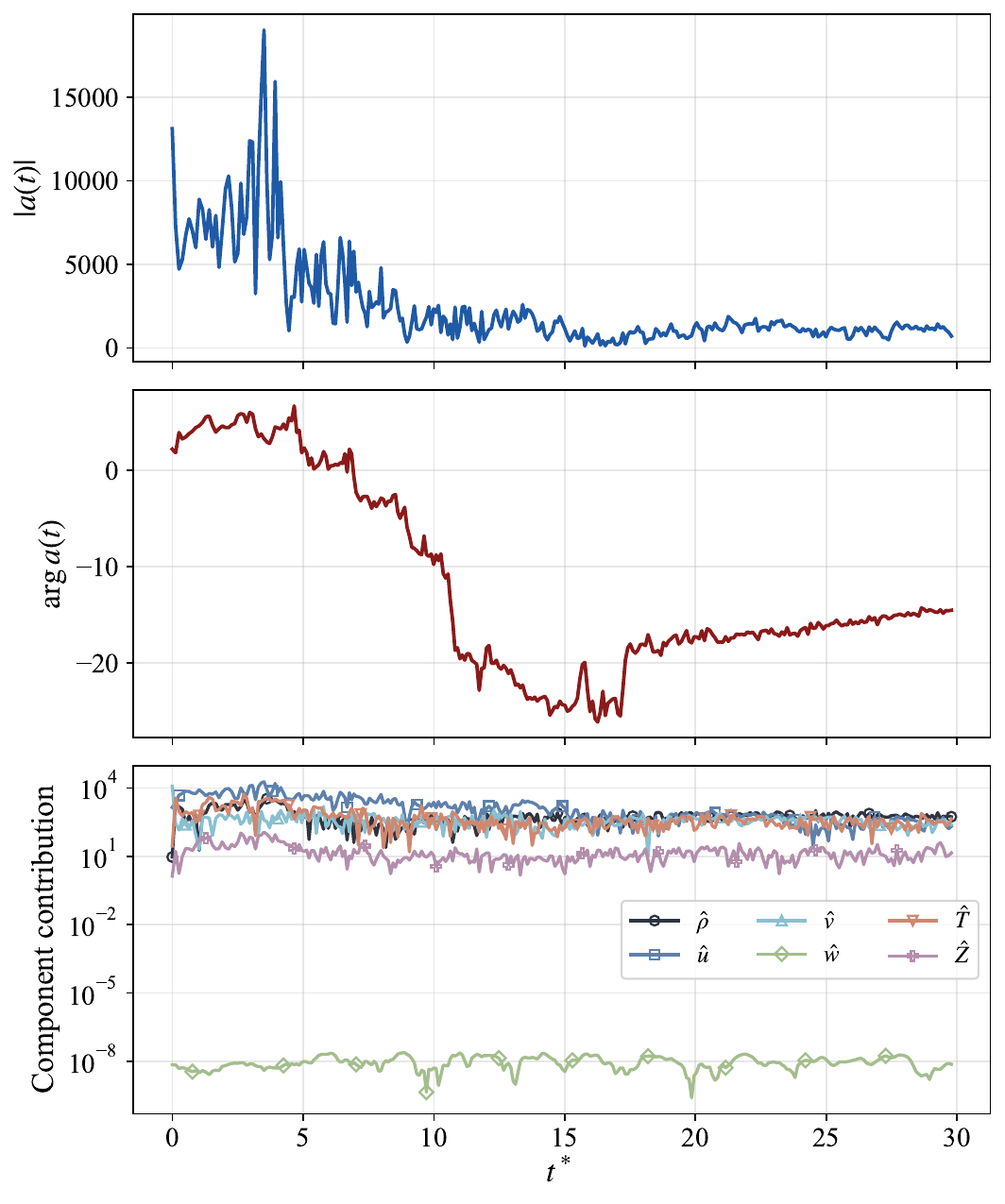}
    \caption{Single-mode amplitude, phase, and componentwise contributions.}
  \end{subfigure}\hfill
  \begin{subfigure}[t]{0.54\textwidth}
    \centering
    \includegraphics[width=\textwidth]{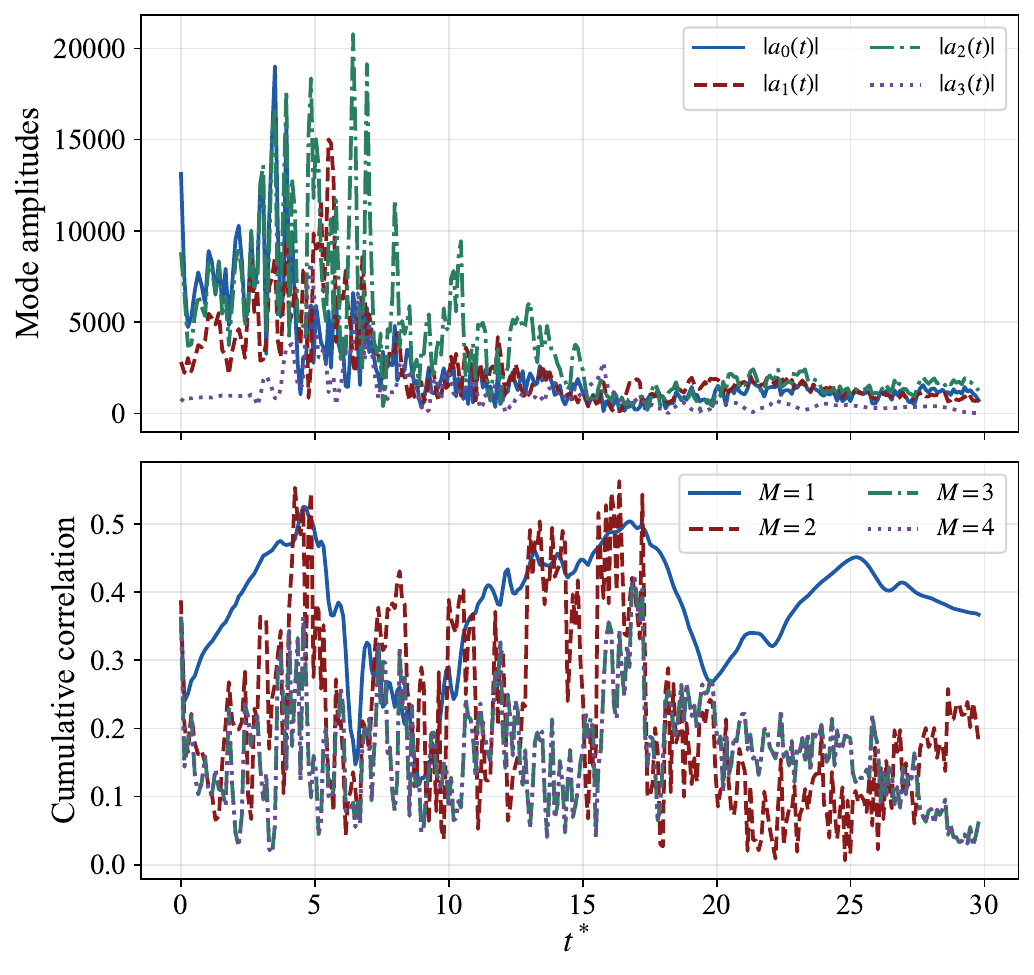}
    \caption{Few-mode biorthogonal decomposition.~\\~}
  \end{subfigure}\\[0.6em]
  \begin{subfigure}[t]{0.58\textwidth}
    \centering
    \includegraphics[width=\textwidth]{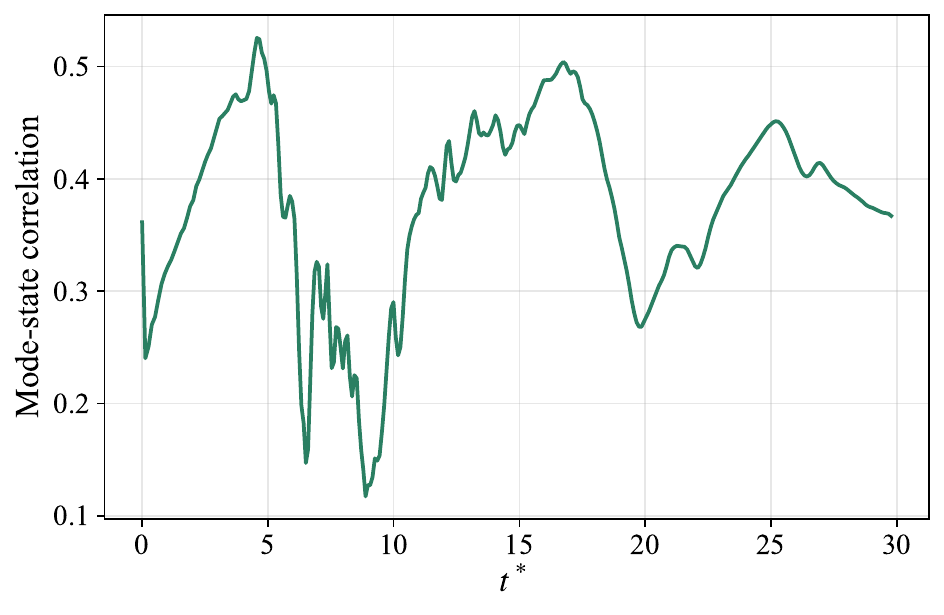}
    \caption{Energy-inner-product correlation with the selected mode.}
  \end{subfigure}
  \caption{Time-resolved modal amplitudes and biorthogonal decomposition at the fundamental streamwise wavenumber. All histories are plotted against the nondimensional time $t^*=tU_{c,0}/\delta_{\omega,0}$.}
  \label{fig:modal_projection}
\end{figure}

\subsection{Comparison with the vortex-sheet reference and prior theory}
The compressible vortex-sheet comparison provides a compact theoretical reference constructed from the outer-stream states of the same reacting temporal layer. Those states are already strongly asymmetric: $U_t\approx \SI{2209}{m/s}$, $U_b\approx \SI{-744}{m/s}$, and the convective reference speed extracted from the two streams is only $U_{\mathrm{conv}}\approx \SI{63.4}{m/s}$. The corresponding convective Mach numbers are high, $M_{c,t}\approx 2.75$ and $M_{c,b}\approx 1.72$, so the vortex-sheet model represents a strongly compressible discontinuous limit of the same mean layer. The reference itself is the classical compressible vortex-sheet problem
\citep{Miles1958}, augmented here by the compressible free-shear-layer
stability context established by Blumen \citep{Blumen1970}, Blumen et al. \citep{BlumenDrazinBillings1975}, Drazin and Davey
\citep{DrazinDavey1977}, and Jackson and Grosch
\citep{JacksonGrosch1989}. Its use here as a receptivity/stability foil follows
the same spirit as the compressible mixing-layer comparisons developed by Barone and Lele \citep{BaroneLele2005}. The corresponding nondimensional outer-state values are collected in Table~\ref{tab:theory_states}.

In that limit, the theoretical branch is essentially neutrally stable over the sampled $k_x^*$ range, with nondimensional growth rates bounded between $-3.96\times 10^{-26}$ and $7.64\times 10^{-18}$ and nondimensional phase speed nearly constant at $c_r^*\approx 0.0817$. The selected numerical branch differs markedly from this reference. At the fundamental wavenumber $k_x^*\approx 0.569$, the numerical operator predicts a slightly damped selected mode with $\sigma^*\approx -9.47\times 10^{-4}$, whereas the same selected family reaches $\sigma^*\approx 7.35\times 10^{-3}$ nearby at $k_x^*\approx 0.627$. The numerical phase speed at the fundamental, $c_r^*\approx 1.80\times 10^{-2}$, also differs strongly from the theoretical $0.0817$. This discrepancy is itself one of the main results. The outer-state compressible jump conditions by themselves predict a stabilized discontinuous branch, whereas the finite-thickness reacting operator supports unstable Kelvin--Helmholtz intervals with much smaller phase speed. The distributed mean shear, together with its reacting thermodynamic and transport structure, therefore reorganizes the classical compressible limit in a dynamically essential way. This comparison sharpens earlier observations from compressible mixing-layer theory and simulation that finite-thickness effects preserve unstable shear-layer structure more robustly than the simplest outer-state arguments alone would suggest
\citep{SandhamReynolds1990,JacksonGrosch1989,FreundLeleMoin2000}. Relative to
the inviscid reacting flame-sheet calculations of Jackson and Grosch
\citep{JacksonGrosch1990},
the present result points to the same broad conclusion: reacting thermodynamic structure can alter instability behaviour qualitatively as well as quantitatively. Here that conclusion is obtained for a finite-thickness mean state and then connected directly to receptivity and modal content in the turbulent reacting layer.
\begin{figure}
  \centering
  \includegraphics[width=0.64\textwidth]{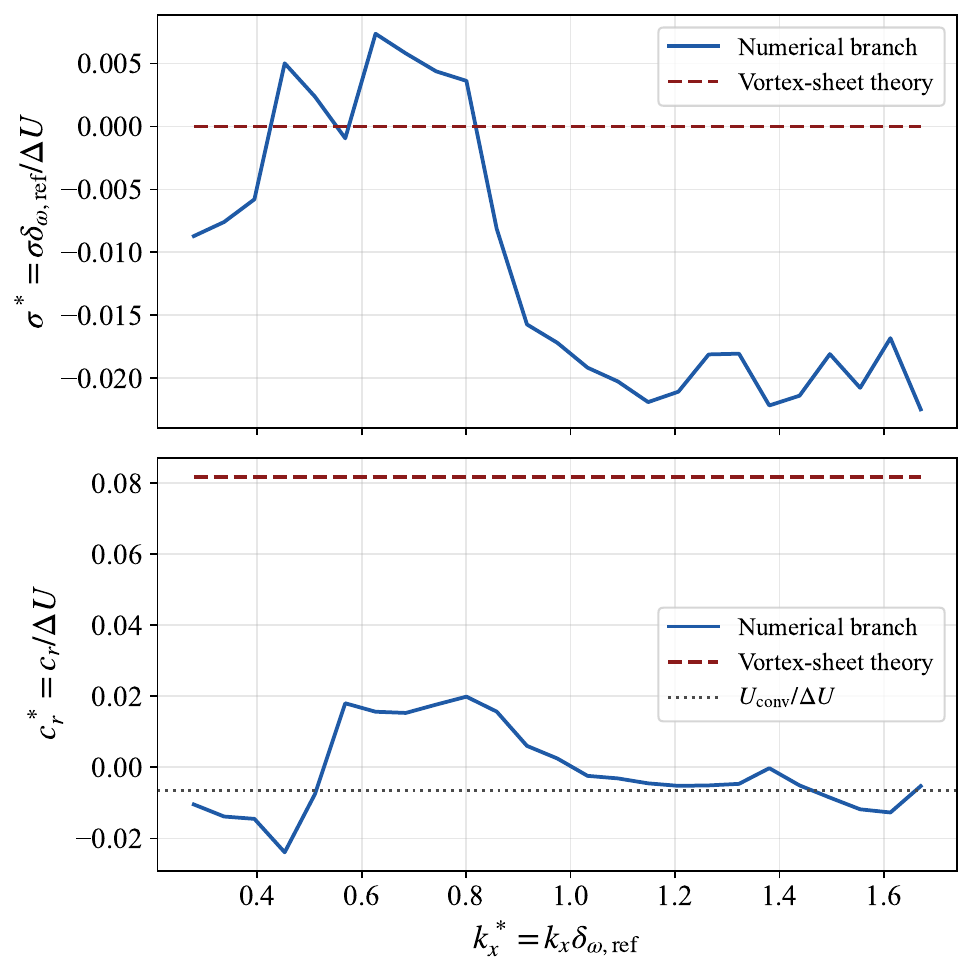}
  \caption{Comparison of the selected numerical Kelvin--Helmholtz branch against the compressible vortex-sheet reference constructed from the outer-stream states of the reacting temporal layer.}
  \label{fig:theory_compare}
\end{figure}

\begin{table}[tbp]
  \centering
  \captionsetup{skip=2pt}
  \caption{Outer-stream states and derived convective parameters used to define the compressible vortex-sheet reference.}
  \label{tab:theory_states}
  \begingroup
  \scriptsize
  \setlength{\tabcolsep}{6pt}
  \renewcommand{\arraystretch}{1.18}
  \rowcolors{2}{tablerow}{white}
  \begin{tabular}{@{}lcc@{}}
    \toprule
    \rowcolor{tablehead}
    \multicolumn{1}{@{}l}{\textbf{Quantity}} &
    \multicolumn{1}{c}{\textbf{Lower stream}} &
    \multicolumn{1}{c@{}}{\textbf{Upper stream}} \\
    \midrule
    $\rho^*=\rho/\rho_b$ & 1.0000 & 0.3244 \\
    $U^*=U/\Delta U$ & -0.2516 & 0.7484 \\
    $T^*=T/T_b$ & 1.0000 & 3.0228 \\
    $Z$ & 1.0000 & 0.0000 \\
    $a^*=a/\Delta U$ & 0.1570 & 0.2537 \\
    $M_c$ & 1.5597 & 2.9757 \\
    \midrule
    \multicolumn{1}{@{}l}{$U_{\mathrm{conv}}^*=U_{\mathrm{conv}}/\Delta U$} & \multicolumn{2}{c@{}}{-0.0066} \\
    \multicolumn{1}{@{}l}{$U_{c,0}^*=U_{c,0}/\Delta U$} & \multicolumn{2}{c@{}}{0.2484} \\
    \addlinespace[2pt]
    \bottomrule
  \end{tabular}
  \endgroup
\end{table}

Taken together, the results establish a consistent hierarchy. The reacting mean state defines a finite-thickness cross-stream structure with substantial thermodynamic and transport variation, the selected Kelvin--Helmholtz branch contains unstable low-to-moderate wavenumber intervals, the direct/adjoint pairing is sufficiently well conditioned to support a receptivity interpretation, and the resulting forcing hierarchy separates raw mass and mixture-fraction sensitivity from heat-release-supported thermal sensitivity. The DNS projections then show that the same branch remains present in the nonlinear planar dynamics, although it does not form a closed one-mode description. The comparison with the discontinuous compressible limit closes the argument by showing that these features are consequences of the distributed reacting shear layer itself, with outer-stream compressibility alone providing only part of the explanation.

\section{Conclusions}
The reacting temporal mixing layer considered here supports a coherent finite-thickness Kelvin--Helmholtz family that can be selected by anchoring at the DNS fundamental wavenumber, paired with its adjoint under an energy-weighted inner product, and used as the organizing structure for both receptivity analysis and biorthogonal modal projection. The mean layer itself is strongly asymmetric in density, velocity, temperature, and composition, and the corresponding thermodynamic and transport closures vary substantially across the layer. The modal problem is therefore controlled by a reacting-base-state variable-coefficient operator, with constant-property reductions unable to capture the same dynamics.

Within this framework, the direct/adjoint pairing is numerically robust, with energy-weighted overlaps close to unity and small biorthogonality errors. The receptivity maps then show a clear ordering of forcing channels. Raw localized mass forcing gives the largest projection onto the selected branch, mixture-fraction forcing is the next-largest raw channel through composition-pressure coupling, and the momentum and thermal channels are weaker in the unweighted maps. The chemistry-weighted thermal map adds the thermochemical-support information: the regions of strongest heat-release support also coincide with large modal sensitivity, making thermal forcing the leading reacting pathway after the source support is included.

The time-resolved planar projections place these receptivity results in contact with the nonlinear DNS data. The selected Kelvin--Helmholtz branch maintains a persistent modal amplitude over the full sampled interval and captures a substantial fraction of the projected fundamental content, but it does not close the dynamics as a single-mode description. The amplitude budget contains substantial streamwise-velocity, density, temperature, and cross-stream-velocity contributions, consistent with a coupled kinematic--thermochemical branch rather than a purely thermal response. The few-mode reconstructions confirm that additional retained modes contribute meaningfully, while also showing that the truncated representation must be interpreted biorthogonally instead of as an orthogonal energy partition.

The compressible vortex-sheet comparison supplies a compact theoretical reference built from the same outer-stream states. In that discontinuous limit, the branch is essentially neutral and propagates at nearly constant phase speed. The finite-thickness reacting operator, by contrast, supports a sustained unstable Kelvin--Helmholtz branch with much smaller phase speed. The gap between these two descriptions shows that the distributed reacting shear layer does more than perturb the classical compressible instability picture; it changes the instability character in a dynamically important way. This observation provides the central platform for interpreting the reacting receptivity maps, the DNS-based modal amplitudes, and the comparison with prior theoretical and numerical studies. More broadly, it indicates that the instability pathways available to turbulent compressible reacting flows cannot be inferred from outer-stream compressibility alone. The distributed thermodynamic state of the reacting layer, and the forcing hierarchy it induces, must also enter any reduced-order interpretation of how turbulent reacting shear layers select and sustain coherent unsteadiness.

The main implication of the present study is physical as much as it is methodological. The finite-thickness thermodynamic structure of the layer reshapes the instability, reorganizes the forcing channels through which it is most readily excited, and leaves a measurable imprint on the nonlinear turbulent dynamics. Any reduced description of turbulent compressible reacting mixing layers that seeks to predict coherent oscillation, sensitivity to forcing, or dominant transport pathways must therefore resolve that coupled shear--thermochemistry structure explicitly or account for it in a model-consistent way.

\clearpage

\appendix
\renewcommand{\theHsection}{appendix.\Alph{section}}

\section{Branch-selection and spanwise-representativeness checks}
\label{app:support_checks}
Two support checks were carried out for the selected branch and the planar projection. The first repeats the branch selection using alternative rules applied to the same eigensweep. Shape-aware tracking anchored at the DNS fundamental, shape-aware tracking anchored at the maximum-growth member of the selected family, and the branch used in the main text give the same fundamental-wavenumber value $\sigma^*\approx -9.47\times 10^{-4}$, the same maximum growth $\sigma^*\approx 7.35\times 10^{-3}$ at $k_x^*\approx 0.627$, and the same unstable intervals $0.453\lesssim k_x^*\lesssim 0.511$ and $0.627\lesssim k_x^*\lesssim 0.801$. Eigenvalue-only tracking gives the same fundamental value and maximum growth but loses the lower-$k_x$ unstable segment, while a local leading-growth selection jumps among additional high-$k_x$ candidates. Across all of these variants, the largest raw channels remain mass and mixture-fraction forcing, and the chemistry-weighted thermal map remains the leading support-weighted thermochemical pathway.

The second check compares the planar modal projection with the full spanwise set of $x$--$y$ planes from the post-developing three-dimensional snapshots, $21.6\lesssim t^*\lesssim 29.8$. For each three-dimensional snapshot, the same streamwise Fourier coefficient, direct/adjoint mode pair, and energy-weighted inner product used in the planar analysis are applied to all 64 spanwise planes. Over the combined spanwise ensemble, the 5th, 50th, and 95th percentiles are approximately $(643,1059,1554)$ for $|a_0|$ and $(0.339,0.394,0.456)$ for $\mathcal{C}^{(1)}$. The planar-slice amplitude samples both high and low parts of the spanwise distribution, while the mode--state correlation remains within the spanwise envelope at the same snapshots. The $x$-mean profiles are also close to the corresponding $x$--$z$ means: the 95th-percentile relative $L_2$ profile differences for $\rho$, $u$, $T$, and $Z$ are all below $1.8\%$. These checks support the use of the planar data as a representative measure of the selected branch's modal imprint, with the absolute amplitude retaining the expected spanwise variability of the turbulent field.

\begin{figure}
  \centering
  \begin{subfigure}[t]{0.82\textwidth}
    \centering
    \includegraphics[width=\textwidth]{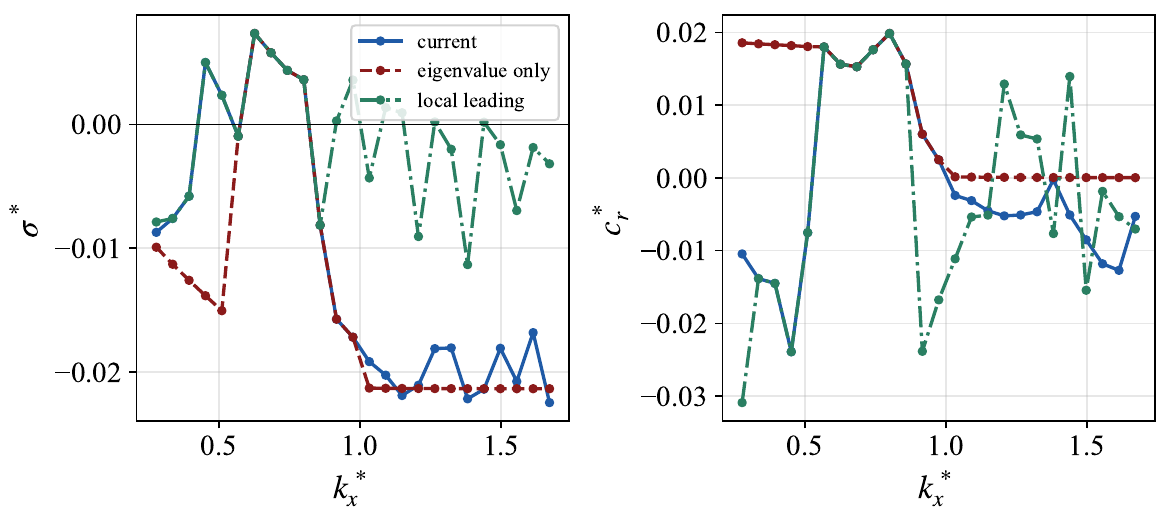}
    \caption{Branch-selection sensitivity.}
  \end{subfigure}\\[0.7em]
  \begin{subfigure}[t]{0.82\textwidth}
    \centering
    \includegraphics[width=\textwidth]{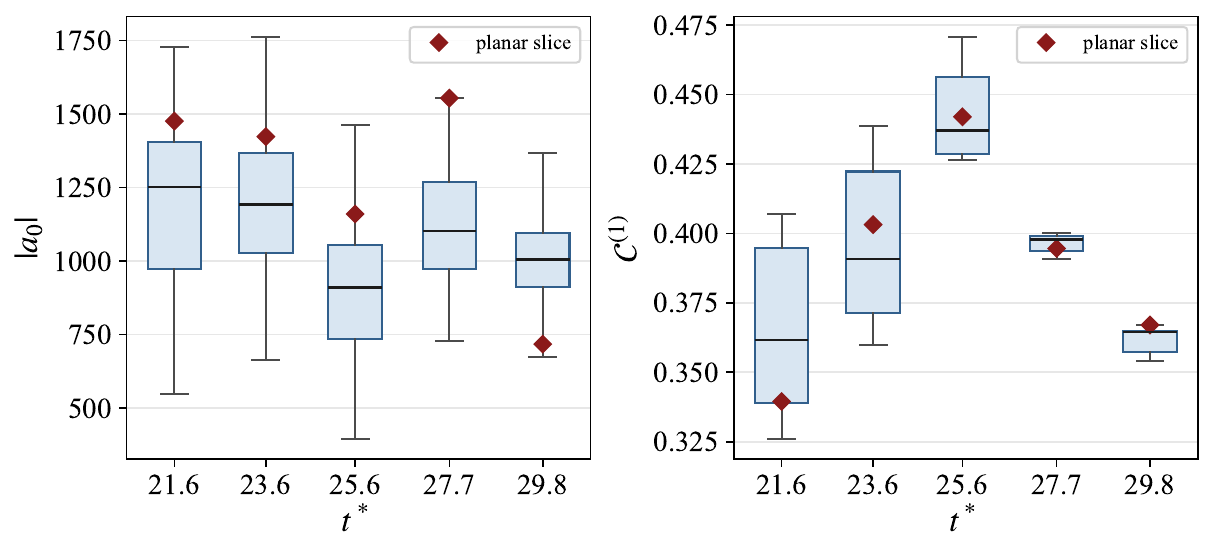}
    \caption{Spanwise representativeness of the planar modal projection.}
  \end{subfigure}
  \caption{Support checks for the selected Kelvin--Helmholtz branch and the planar modal projection. In (a), the shape-aware fundamental-anchor and peak-growth-anchor alternatives collapse onto the branch used in the main analysis; the plotted alternatives show the effect of using eigenvalue-only tracking or local leading-growth selection. In (b), boxes show the distribution over all 64 spanwise planes from each three-dimensional snapshot, and diamonds mark the planar slice used in the main time-resolved projection.}
  \label{fig:support_checks}
\end{figure}

\section{Additional verification diagnostics}
Two additional numerical checks are retained here for completeness. The first evaluates the discrete weighted Lagrange-identity defect associated with the energy inner product used to define the adjoint,
\begin{equation}
\delta_{\mathcal{L}}(k_x)=
\frac{\left|\langle a,\mathcal{L}b\rangle_E-\langle \mathcal{L}^\dagger a,b\rangle_E\right|}
{\max\!\left(\left|\langle a,\mathcal{L}b\rangle_E\right|,\left|\langle \mathcal{L}^\dagger a,b\rangle_E\right|,\epsilon\right)},
\end{equation}
where $\epsilon$ is a small positive floor introduced only to avoid division by zero in degenerate cases. Over the full $k_x$ sweep, the relative defect remains between \num{2.31e-15} and \num{1.85e-14}. The second reports the left--right biorthogonality matrix
\begin{equation}
B_{ij}=p_i^*q_j
\end{equation}
at the fundamental wavenumber. Its diagonal entries remain at unity to roundoff, while the largest off-diagonal entry is only \num{2.15e-6}. These checks are more algebraic than the residual and forced-response checks reported in the main text, but they further confirm that the discrete direct/adjoint system is internally consistent.

\begin{figure}[htbp]
  \centering
  \begin{subfigure}{0.37\textwidth}
    \centering
    \includegraphics[width=\textwidth]{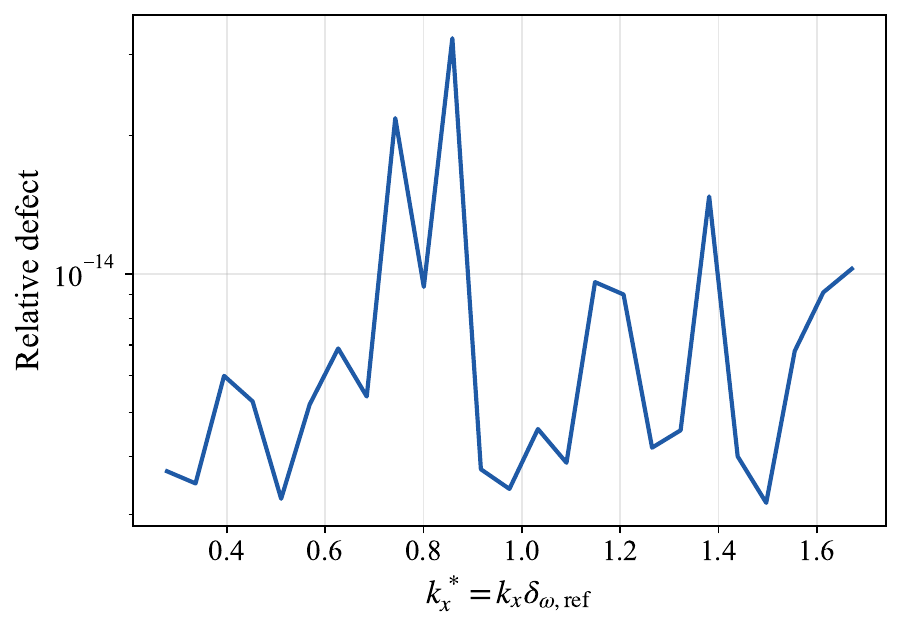}
    \caption{Discrete weighted Lagrange-identity defect.}
  \end{subfigure}\hfill
  \begin{subfigure}{0.59\textwidth}
    \centering
    \includegraphics[width=\textwidth]{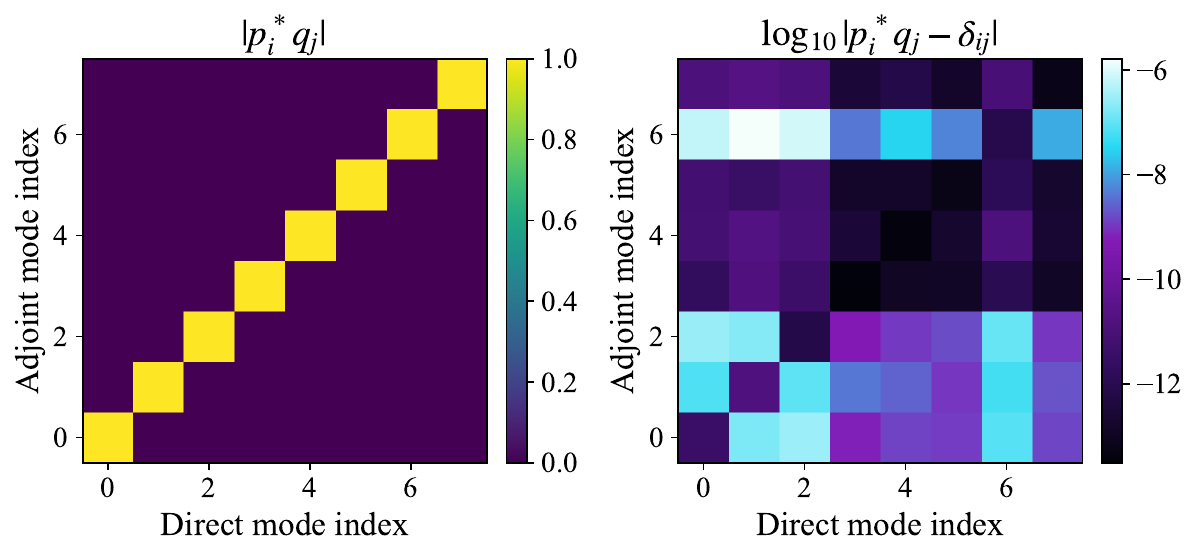}
    \caption{Left--right biorthogonality matrix at the fundamental.}
  \end{subfigure}
  \caption{Supplementary verification diagnostics retained for completeness. Both checks support the numerical consistency of the direct/adjoint formulation used in the main text.}
  \label{fig:appendix_verification}
\end{figure}

\clearpage

\section*{Acknowledgements}
The computing power for this study was provided by the Phoenix Computing Cluster as a part of Georgia Tech's Partnership for Advanced Computing Environment, and is gratefully acknowledged.

\section*{Funding}
This research received no specific grant from any funding agency, commercial or not-for-profit sectors.

\section*{Author contributions}
Sriram P. Kalathoor: Conceptualization, Methodology, Software, Formal analysis, Investigation, Visualization, Writing -- original draft. Joseph C. Oefelein: Conceptualization, Resources, Supervision, Writing -- review and editing.

\section*{Competing interests}
The authors have no competing interests to declare.

\section*{Data availability}
The data that support the findings of this study are available from the corresponding author upon reasonable request.

\bibliographystyle{spmpsci}
\bibliography{references}

\begin{thebibliography}{10}
\providecommand{\url}[1]{{#1}}
\providecommand{\urlprefix}{URL }
\expandafter\ifx\csname urlstyle\endcsname\relax
  \providecommand{\doi}[1]{DOI~\discretionary{}{}{}#1}\else
  \providecommand{\doi}{DOI~\discretionary{}{}{}\begingroup
  \urlstyle{rm}\Url}\fi

\bibitem{AlHasnineRussoTuminBrehm2023}
Al~Hasnine, S.M.A., Russo, V., Tumin, A., Brehm, C.: Biorthogonal decomposition
  of the disturbance flow field generated by particle impingement on a
  hypersonic boundary layer.
\newblock Journal of Fluid Mechanics \textbf{969}, A1 (2023).
\newblock \doi{10.1017/jfm.2023.531}

\bibitem{Balsa1988}
Balsa, T.F.: On the receptivity of free shear layers to two-dimensional
  external excitation.
\newblock Journal of Fluid Mechanics \textbf{187}, 155--177 (1988).
\newblock \doi{10.1017/S0022112088000382}

\bibitem{BaroneLele2005}
Barone, M.F., Lele, S.K.: Receptivity of the compressible mixing layer.
\newblock Journal of Fluid Mechanics \textbf{540}, 301--335 (2005).
\newblock \doi{10.1017/S0022112005005884}

\bibitem{Bellan2017}
Bellan, J.: Direct numerical simulation of a high-pressure turbulent reacting
  temporal mixing layer.
\newblock Combustion and Flame \textbf{176}, 245--262 (2017).
\newblock \doi{10.1016/j.combustflame.2016.09.026}

\bibitem{Blumen1970}
Blumen, W.: Shear layer instability of an inviscid compressible fluid.
\newblock Journal of Fluid Mechanics \textbf{40}(4), 769--781 (1970).
\newblock \doi{10.1017/S0022112070000435}

\bibitem{BlumenDrazinBillings1975}
Blumen, W., Drazin, P.G., Billings, D.F.: Shear layer instability of an
  inviscid compressible fluid. {P}art 2.
\newblock Journal of Fluid Mechanics \textbf{71}(2), 305--316 (1975).
\newblock \doi{10.1017/S0022112075002595}

\bibitem{Breidenthal1981}
Breidenthal, R.E.: Structure in turbulent mixing layers and wakes using a
  chemical reaction.
\newblock Journal of Fluid Mechanics \textbf{109}, 1--24 (1981).
\newblock \doi{10.1017/S0022112081000906}

\bibitem{BroadwellBreidenthal1982}
Broadwell, J.E., Breidenthal, R.E.: A simple model of mixing and chemical
  reaction in a turbulent shear layer.
\newblock Journal of Fluid Mechanics \textbf{125}, 397--410 (1982).
\newblock \doi{10.1017/S0022112082003401}

\bibitem{BrownRoshko1974}
Brown, G.L., Roshko, A.: On density effects and large structure in turbulent
  mixing layers.
\newblock Journal of Fluid Mechanics \textbf{64}(4), 775--816 (1974).
\newblock \doi{10.1017/S002211207400190X}

\bibitem{Chu1965}
Chu, B.T.: On the energy transfer to small disturbances in fluid flow ({P}art
  {I}).
\newblock Acta Mechanica \textbf{1}, 215--234 (1965).
\newblock \doi{10.1007/BF01387235}

\bibitem{ClemensMungal1995}
Clemens, N.T., Mungal, M.G.: Large-scale structure and entrainment in the
  supersonic mixing layer.
\newblock Journal of Fluid Mechanics \textbf{284}, 171--216 (1995).
\newblock \doi{10.1017/S0022112095000358}

\bibitem{DimotakisBrown1976}
Dimotakis, P.E., Brown, G.L.: The mixing layer at high {R}eynolds number:
  large-structure dynamics and entrainment.
\newblock Journal of Fluid Mechanics \textbf{78}(3), 535--560 (1976).
\newblock \doi{10.1017/S0022112076002590}

\bibitem{DrazinDavey1977}
Drazin, P.G., Davey, A.: Shear layer instability of an inviscid compressible
  fluid. {P}art 3.
\newblock Journal of Fluid Mechanics \textbf{82}(2), 255--283 (1977).
\newblock \doi{10.1017/S0022112077000640}

\bibitem{ElliottSamimy1990}
Elliott, G.S., Samimy, M.: Compressibility effects in free shear layers.
\newblock Physics of Fluids A \textbf{2}(7), 1231--1240 (1990).
\newblock \doi{10.1063/1.857816}

\bibitem{FoysiSarkar2010}
Foysi, H., Sarkar, S.: The compressible mixing layer: an {LES} study.
\newblock Theoretical and Computational Fluid Dynamics \textbf{24}(6), 565--588
  (2010).
\newblock \doi{10.1007/s00162-009-0176-8}

\bibitem{FreundLeleMoin2000}
Freund, J.B., Lele, S.K., Moin, P.: Compressibility effects in a turbulent
  annular mixing layer. {P}art 1. turbulence and growth rate.
\newblock Journal of Fluid Mechanics \textbf{421}, 229--267 (2000).
\newblock \doi{10.1017/S0022112000001622}

\bibitem{Goldstein1983}
Goldstein, M.E.: The evolution of {T}ollmien--{S}chlichting waves near a
  leading edge.
\newblock Journal of Fluid Mechanics \textbf{127}, 59--81 (1983).
\newblock \doi{10.1017/S002211208300261X}

\bibitem{Goldstein1985}
Goldstein, M.E.: Scattering of acoustic waves into {T}ollmien--{S}chlichting
  waves by small streamwise variations in surface geometry.
\newblock Journal of Fluid Mechanics \textbf{154}, 509--529 (1985).
\newblock \doi{10.1017/S0022112085001641}

\bibitem{HermansonDimotakis1989}
Hermanson, J.C., Dimotakis, P.E.: Effects of heat release in a turbulent,
  reacting shear layer.
\newblock Journal of Fluid Mechanics \textbf{199}, 333--375 (1989).
\newblock \doi{10.1017/S0022112089000406}

\bibitem{Hill1995}
Hill, D.C.: Adjoint systems and their role in the receptivity problem for
  boundary layers.
\newblock Journal of Fluid Mechanics \textbf{292}, 183--204 (1995).
\newblock \doi{10.1017/S0022112095001480}

\bibitem{JacksonGrosch1989}
Jackson, T.L., Grosch, C.E.: Inviscid spatial stability of a compressible
  mixing layer.
\newblock Journal of Fluid Mechanics \textbf{208}, 609--637 (1989).
\newblock \doi{10.1017/S002211208900296X}

\bibitem{JacksonGrosch1990}
Jackson, T.L., Grosch, C.E.: Inviscid spatial stability of a compressible
  mixing layer. part 2. the flame sheet model.
\newblock Journal of Fluid Mechanics \textbf{217}, 391--420 (1990).
\newblock \doi{10.1017/S0022112090000775}

\bibitem{JahanbakhshiMadnia2016}
Jahanbakhshi, R., Madnia, C.K.: Entrainment in a compressible turbulent shear
  layer.
\newblock Journal of Fluid Mechanics \textbf{797}, 564--603 (2016).
\newblock \doi{10.1017/jfm.2016.296}

\bibitem{JahanbakhshiMadnia2018}
Jahanbakhshi, R., Madnia, C.K.: The effect of heat release on the entrainment
  in a turbulent mixing layer.
\newblock Journal of Fluid Mechanics \textbf{844}, 92--126 (2018).
\newblock \doi{10.1017/jfm.2018.122}

\bibitem{KnausPantano2009}
Knaus, R.C., Pantano, C.: On the effect of heat release in turbulence spectra
  of non-premixed reacting shear layers.
\newblock Journal of Fluid Mechanics \textbf{626}, 67--109 (2009).
\newblock \doi{10.1017/S0022112008005636}

\bibitem{MahleEtAl2007}
Mahle, S., Dowling, A.P., Luo, K.H., Sandham, N.D.: On the turbulence structure
  in inert and reacting compressible mixing layers.
\newblock Journal of Fluid Mechanics \textbf{593}, 171--180 (2007).
\newblock \doi{10.1017/S0022112007008919}

\bibitem{MatsunoLele2020}
Matsuno, K., Lele, S.K.: Internal regulation in compressible turbulent shear
  layers.
\newblock Journal of Fluid Mechanics \textbf{900}, A5 (2020).
\newblock \doi{10.1017/jfm.2020.492}

\bibitem{McmurtryRileyMetcalfe1989}
McMurtry, P.A., Riley, J.J., Metcalfe, R.W.: Effects of heat release on the
  large-scale structure in turbulent mixing layers.
\newblock Journal of Fluid Mechanics \textbf{199}, 297--332 (1989).
\newblock \doi{10.1017/S002211208900039X}

\bibitem{Miles1958}
Miles, J.W.: On the disturbed motion of a plane vortex sheet.
\newblock Journal of Fluid Mechanics \textbf{4}(5), 538--552 (1958).
\newblock \doi{10.1017/S0022112058000653}

\bibitem{PantanoSarkarWilliams2002}
Pantano, C., Sarkar, S., Williams, F.A.: The interaction of scalar mixing and
  heat release in a reacting shear layer.
\newblock In: Turbulent Mixing and Beyond, pp. 137--148. Springer, Dordrecht
  (2002).
\newblock \doi{10.1007/978-94-017-1998-8_11}

\bibitem{PantanoSarkarWilliams2003}
Pantano, C., Sarkar, S., Williams, F.A.: Mixing of a conserved scalar in a
  turbulent reacting shear layer.
\newblock Journal of Fluid Mechanics \textbf{481}, 291--328 (2003).
\newblock \doi{10.1017/S0022112003003872}

\bibitem{PapamoschouRoshko1988}
Papamoschou, D., Roshko, A.: The compressible turbulent shear layer: an
  experimental study.
\newblock Journal of Fluid Mechanics \textbf{197}, 453--477 (1988).
\newblock \doi{10.1017/S0022112088003325}

\bibitem{SandhamReynolds1990}
Sandham, N.D., Reynolds, W.C.: Compressible mixing layer: linear theory and
  direct simulation.
\newblock AIAA Journal \textbf{28}(4), 618--624 (1990).
\newblock \doi{10.2514/3.10437}

\bibitem{SandhamReynolds1991}
Sandham, N.D., Reynolds, W.C.: Three-dimensional simulations of large eddies in
  the compressible mixing layer.
\newblock Journal of Fluid Mechanics \textbf{224}, 133--158 (1991).
\newblock \doi{10.1017/S0022112091001719}

\bibitem{Sarkar1995}
Sarkar, S.: The stabilizing effect of compressibility in turbulent shear flow.
\newblock Journal of Fluid Mechanics \textbf{282}, 163--186 (1995).
\newblock \doi{10.1017/S0022112095000134}

\bibitem{Schmid2007}
Schmid, P.J.: Nonmodal stability theory.
\newblock Annual Review of Fluid Mechanics \textbf{39}, 129--162 (2007).
\newblock \doi{10.1146/annurev.fluid.38.050304.092139}

\bibitem{TairaEtAl2017}
Taira, K., Hemati, M.S., Brunton, S.L., Sun, Y., Duraisamy, K., Bagheri, S.,
  Dawson, S.T.M., Yeh, C.A.: Modal analysis of fluid flows: an overview.
\newblock AIAA Journal \textbf{55}(12), 4013--4041 (2017).
\newblock \doi{10.2514/1.J056060}

\bibitem{Tam1978}
Tam, C.K.W.: Excitation of instability waves in a two-dimensional shear layer
  by sound.
\newblock Journal of Fluid Mechanics \textbf{89}(2), 357--371 (1978).
\newblock \doi{10.1017/S0022112078002674}

\bibitem{TuminWangZhong2007}
Tumin, A., Wang, X., Zhong, X.: Direct numerical simulation and the theory of
  receptivity in a hypersonic boundary layer.
\newblock Physics of Fluids \textbf{19}(1), 014,101 (2007).
\newblock \doi{10.1063/1.2409731}

\bibitem{VremanSandhamLuo1996}
Vreman, A.W., Sandham, N.D., Luo, K.H.: Compressible mixing layer growth rate
  and turbulence characteristics.
\newblock Journal of Fluid Mechanics \textbf{320}, 235--258 (1996).
\newblock \doi{10.1017/S0022112096007525}

\bibitem{WuTumin2019}
Wu, L., Tumin, A.: Receptivity of turbulent compressible mixing layers to
  localized energy deposition.
\newblock In: AIAA Scitech 2019 Forum, pp. 2019--1652 (2019).
\newblock \doi{10.2514/6.2019-1652}

\end{thebibliography}

\end{document}